\documentclass[12pt]{article}
\pdfoutput=1
\usepackage{latexsym}
\usepackage{amsmath,,calrsfs}
\usepackage{amsfonts}
\usepackage{amssymb}
\usepackage{amscd}
\usepackage{bbm}
\usepackage{fancybox}
\usepackage{cite}
\usepackage{amsmath,amsfonts,amsbsy}
\usepackage{pstricks,pst-node}
\usepackage[small,bf,hang]{caption2}
\usepackage{psfrag}
\usepackage{comment}

\usepackage{multirow}                     
\usepackage{float}                          
\usepackage{lscape}                         
\usepackage{bm}

\newcommand{\bb}{\bar\beta}

\newcommand{\beq}{\begin{equation}}
\newcommand{\eeq}{\end{equation}}
\newcommand{\bt}{\begin{tabular}}
\newcommand{\et}{\end{tabular}}
\newcommand{\bc}{\begin{center}}
\newcommand{\ec}{\end{center}}

\newcommand{\be}{\begin{equation}}
\newcommand{\ee}{\end{equation}}
\newcommand{\bea}{\begin{eqnarray}}
\newcommand{\eea}{\end{eqnarray}}
\newcommand{\ba}{\begin{array}}
\newcommand{\ea}{\end{array}}

\def\bbox{{\,\lower0.9pt\vbox{\hrule \hbox{\vrule height 0.2 cm
\hskip 0.2 cm \vrule height 0.2 cm}\hrule}\,}}
\newcommand{\dsl}{\pa \kern-0.5em /}

\font\mybb=msbm10 at 12pt
\def\bb#1{\hbox{\mybb#1}}

\def\bR {\bb{R}}

\def\BB{{\bf B}}
\def\EE{{\bf E}}
\def\DD{{\bf D}}

\makeatletter \@addtoreset{equation}{section} \makeatother

\def\slashchar#1{\setbox0=\hbox{$#1$}           
   \dimen0=\wd0                                 
   \setbox1=\hbox{/} \dimen1=\wd1               
   \ifdim\dimen0>\dimen1                        
      \rlap{\hbox to \dimen0{\hfil/\hfil}}      
      #1                                        
   \else                                        
      \rlap{\hbox to \dimen1{\hfil$#1$\hfil}}   
      /                                         
   \fi}

\date{}

\begin{document}

\begin{titlepage}

\begin{center}

\vskip 1.5cm

{\Large \bf  Hamiltonian  birefringence and Born-Infeld limits}
\smallskip

\vskip 1cm

{\bf Luca Mezincescu\,${}^{a}$, Jorge G.~Russo\,${}^{b,c}$ and  Paul K.~Townsend\,${}^d$} \\

\vskip 25pt

{\em $^a$  \hskip -.1truecm
\em Department of Physics, University of Miami, P.O. Box 248046, \\
Coral Gables, FL 33124, USA
 \vskip 5pt }

\vskip .4truecm

{\em $^b$  \hskip -.1truecm
\em Instituci\'o Catalana de Recerca i Estudis Avan\c{c}ats (ICREA),\\
Pg. Lluis Companys, 23, 08010 Barcelona, Spain.
 \vskip 5pt }

\vskip .4truecm

{\em $^c$  \hskip -.1truecm
\em Departament de F\' \i sica Cu\' antica i Astrof\'\i sica and Institut de Ci\`encies del Cosmos,\\ 
Universitat de Barcelona, Mart\'i Franqu\`es, 1, 08028
Barcelona, Spain.
 \vskip 5pt }
 
 \vskip .4truecm

{\em $^d$ \hskip -.1truecm
\em  Department of Applied Mathematics and Theoretical Physics,\\ Centre for Mathematical Sciences, University of Cambridge,\\
Wilberforce Road, Cambridge, CB3 0WA, U.K.\vskip 5pt }

\hskip 1cm

\noindent {\it e-mail:}  {\texttt  mezincescu@miami.edu, jorge.russo@icrea.cat, pkt10@cam.ac.uk}

\end{center}

\vskip 0.5cm
\begin{center} {\bf ABSTRACT}\\[3ex]
\end{center}

Using Hamiltonian methods, we find six relativistic theories of nonlinear electrodynamics for which plane wave perturbations about a constant uniform background are {\sl not} birefringent. All have the same conformal strong-field limit to Bialynicki-Birula (BB) electrodynamics, but only four avoid superluminal propagation:  Born-Infeld (BI), its non-conformal ``extreme'' limits (electric and magnetic) and the conformal BB limit. The quadratic dispersion relation of BI is shown to degenerate in the extreme limits to a pair of linear relations, which become identical in the BB limit.

\vfill

\end{titlepage}
\tableofcontents

\section{Introduction}

In previous work \cite{Russo:2022qvz} two of us revisited old results of Boillat, Pleba{\'n}ski and Bialynicki-Birula \cite{Boillat:1966eyw,Boillat:1970gw,Plebanski:1970zz,Bialynicki-Birula:1984daz} on birefringence in nonlinear electrodynamics (NLED), which is the class of theories conventionally defined by a Lagrangian density $\mathcal{L}(S,P)$, with
\be
S= \frac12 (|{\bf E}|^2 - |{\bf B}|^2) \, , \qquad P= {\bf E}\cdot{\bf B}\, ,  
\ee
where $({\bf E}, {\bf B})$ are the electric/magnetic field components of a 2-form field strength $F=dA$ on 4D Minkowski spacetime: 
\be
{\bf E} = \boldsymbol{\nabla} A_0 - \dot{\bf A}\, , \qquad {\bf B} = \boldsymbol{\nabla} \times {\bf A}\, . 
\ee
The best-known of these results is that Born-Infeld (BI) theory \cite{Born:1934gh} is the unique zero-birefringence NLED with a weak-field limit; a comparison of the different methods used to obtain this result can be found in \cite{McCarthy:1999hv}. 
Other zero-birefringence cases (without a weak-field limit) were found by Pleba{\'n}ski, and a corollary of the complete catalog presented in \cite{Russo:2022qvz} is that none of them is electromagnetic-duality invariant, in contrast to BI. A further result of \cite{Russo:2022qvz} was that almost all zero-birefringence NLEDs other than Born-Infeld are unphysical because they allow superluminal propagation\footnote{This is presumably related to the earlier finding by other means that duality invariance and ``good propagation'' single out Born-Infeld \cite{Deser:1998wv}.}. 

The qualification ``almost'' arises because the zero-birefringence conditions in a form found by Boillat \cite{Boillat:1970gw} have a solution that yields a Lagrangian constraint rather than a Lagrangian. Imposing this constraint with a Lagrange multiplier provides a Lagrangian density for what was called in \cite{Russo:2022qvz} ``extreme-Born-Infeld'' electrodynamics: 
\be
\mathcal{L}_{\rm eBI} = \lambda \left(T^2 -2TS -P^2\right)\, , 
\ee
where $T$ is the Born-Infeld constant with dimensions of energy density. This takes us outside the class of NLEDs as 
conventionally defined by a Lagrangian density  $\mathcal{L}(S,P)$, but after taking the Legendre transform with respect to 
${\bf E}$ the field $\lambda$ becomes an auxiliary field that can be eliminated by its algebraic field equation to give the following Hamiltonian density
\cite{Russo:2022qvz}:
\be
\mathcal{H}_{\rm eBI} = \sqrt{|{\bf D}\times {\bf B}|^2 + T|{\bf D}|^2}\, , 
\ee
where  the (electric-displacement) field ${\bf D}$ is the Legendre dual to ${\bf E}$. This can be viewed as a particular scaling limit of the BI Hamiltonian density, and a similar scaling limit with the roles of ${\bf D}$ and ${\bf B}$ reversed yields the Hamiltonian density for ``magnetic-extreme-Born-Infeld'' (meBI) electrodynamics:
\be
\mathcal{H}_{\rm meBI} = \sqrt{|{\bf D}\times {\bf B}|^2 + T|{\bf B}|^2}\, .  
\ee
For both eBI and meBI, the strong-field limit (equivalent to $T\to0$ for fixed non-zero energy density) yields 
\be\label{HamBB}
\mathcal{H}_{\rm BB}= \left|{\bf D}\times {\bf B}\right|\, , 
\ee
which is the Hamiltonian density of Bialynicki-Birula (BB) electrodynamics, originally found as the strong-field limit of BI \cite{Bialynicki-Birula:1984daz}.

Although neither meBI nor BB electrodynamics were  found in \cite{Russo:2022qvz} from an analysis of the ``Boillat equations'', a re-examination of the general solution of these equations (eq. (2.40) of \cite{Russo:2022qvz}) shows that all three BI limits are special cases once one allows for Lagrangian constraints. In particular, both meBI and BB arise on the parameter-branch that yields the Pleba{\'n}ski case; on this branch the two Boillat equations (eq. (2.49) of \cite{Russo:2022qvz}) involve two dimensionful parameters. When the ``Pleba{\'n}ski'' parameter ($\kappa$) is non zero we get the Pleba{\'n}ski case; when it is zero we get the constraint $P=0$ from one Boillat equation, while the other equation depends on the second dimensionful parameter ($c-c')$, which may be zero or non-zero. When this second parameter is non-zero we find\footnote{The overall sign is required for $\mathcal{H}\ge0$, and $T=2(c-c')>0$ 
for convexity of $\mathcal{H}$.} $\mathcal{L} = -\sqrt{-2TS}$; this is the ``canonical'' part of the meBI Lagrangian density found in \cite{Russo:2022qvz}. If the second dimensionful parameter is zero we get $S=0$ as an additional constraint; this leads to the BB Lagrangian density of \cite{Bialynicki-Birula:1992rcm}.

However, since Lagrangian densities for the BI limits all require constraints they are not of the initially assumed form $\mathcal{L}(S,P)$, which means that the birefringence status of the BI limits was not settled beyond doubt in \cite{Russo:2022qvz}. The obvious way to resolve this issue is to reconsider birefringence from a Hamiltonian perspective. 

In the case of 6D chiral nonlinear 2-form electrodynamics, where small-amplitude plane wave solutions in a constant `magnetic' background have three independent polarisations, the Hamiltonian formulation was recently used to find and solve the conditions for zero trirefringence (the same dispersion relation for all three polarisation modes) \cite{Bandos:2023yat}. As all 4D NLEDs with an $SO(2)$ electromagnetic duality invariance are the dimensional reduction of some 6D chiral 2-form 
theory \cite{Perry:1996mk,Bandos:2020hgy}, it was possible to deduce from these 6D results that the only duality invariant 4D NLEDs without birefringence are BI and BB. This result is consistent with the unique status of BI in Lagrangian analyses of 4D birefringence because BB electrodynamics, like eBI and meBI, has no `standard' Lagrangian formulation, but it provides no information about eBI and meBI because these are {\sl not}  duality invariant.  What we need therefore is a Hamiltonian analysis of birefringence for 4D NLED that does not assume duality invariance.

A convenient starting point is a `phase-space' Lagrangian density of the form
\be\label{start}
\widetilde{\mathcal{L}}  = {\bf E} \cdot {\bf D} - \mathcal{H}({\bf D},{\bf B})\, , 
\ee
which is a function of $({\bf E}, {\bf B})$ {\sl and} ${\bf D}$. 
The gauge invariant Hamiltonian field equations 
and Bianchi identities are the first-order ``macroscopic Maxwell equations''
\be\label{feqs}
\begin{aligned} 
\dot {\bf D}  &= {}\;\; \boldsymbol{\nabla} \times {\bf H} \, , \qquad  \boldsymbol{\nabla} \cdot {\bf D} =0\\
\dot {\bf B}  &=  -\boldsymbol{\nabla}\times {\bf E} \, , \qquad \boldsymbol{\nabla} \cdot {\bf B} =0 \, , 
\end{aligned}
\ee
which must be taken together with the ``constitutive relations'' 
\begin{equation}\label{EHdefs}
{\bf E} =  \partial {\cal H}/\partial {\bf D} \, , 
\qquad {\bf H} = \partial {\cal H}/\partial {\bf B} \, . 
 \end{equation}
The first of these relations is obtained by variation of ${\bf D}$ in \eqref{start}; the second defines ${\bf H}$. 

As we assume invariance under time and space translations, and space rotations, it must be possible to express 
$\mathcal{H}$ as a function of the three rotation scalars:
\be
x = \frac12|{\bf D}|^2\, , \qquad y= \frac12|{\bf B}|^2 \, , \qquad z= {\bf D}\cdot{\bf B}\, ,  
\ee
in which case the constitutive relations become
\be\label{constit}
{\bf E} = \mathcal{H}_x {\bf D} + \mathcal{H}_z {\bf B} \, , \qquad
{\bf H} = \mathcal{H}_y {\bf B} + \mathcal{H}_z {\bf D} \, .   
\ee
Electromagnetic duality acts on the complex field  $({\bf D} +i{\bf B})$ by a constant shift of its phase, and 
the Hamiltonian density $\mathcal{H}$ is duality invariant if  ${\bf D}\cdot {\bf H}= {\bf E} \cdot {\bf B}$  \cite{Bialynicki-Birula:1984daz}, which is equivalent to 
 \be\label{duality}
 z\left(\mathcal{H}_x-\mathcal{H}_y\right) = 2(x-y)\mathcal{H}_z\, . 
 \ee
However, as emphasised above, we do {\sl not} impose this condition here. We do impose the condition for 
$\mathcal{H}$ to define a Lorentz invariant theory, which is \cite{Bialynicki-Birula:1984daz}
\be\label{BBLI}
{\bf E}\times {\bf H}= c^2 \, {\bf D} \times {\bf B}\, ,  
\ee
where $c$ is the speed of light. We shall set $c=1$, in which case an equivalent condition  is 
\be\label{LI}
\mathcal{I}: = \mathcal{H}_x\mathcal{H}_y -\mathcal{H}_z^2 =1\, .
\ee
This condition ensures the existence of a symmetric stress-energy tensor satisfying the usual continuity conditions. The field momentum density ${\bf p}$ and its magnitude $p$ are given by 
\be
{\bf p} = {\bf D}\times {\bf B}\, , \qquad p^2 =4xy-z^2\, . 
\ee

One solution of the Hamiltonian field equations is 
\be\label{backg}
({\bf D},{\bf B}) = (\bar {\bf D},\bar {\bf B}) \, , 
\ee
where $\bar {\bf D}$ and $\bar {\bf B}$ are arbitrary {\sl constant and uniform} 3-vector densities. By expanding the full 
field equations to first order about such a background we obtain linear equations with plane wave solutions
that propagate in the homogeneous optical medium provided by the background. This medium is ``birefringent'' if the two independent polarizations of the plane waves have different dispersion relations. In general the medium is stationary rather 
than static because its momentum density $\bar{\bf p}$ is non-zero.  Generic stationary backgrounds are Lorentz boosts of static backgrounds, although there may be special cases for which this is not true. 

Of interest here are those special choices of $\mathcal{H}$ for which the optical medium provided by 
any constant uniform electromagnetic background is one {\sl without} birefringence. We restrict the search to 
relativistic NLED, for which the condition $\mathcal{I}=1$ must be satisfied. Our main result is that the Hamiltonian density
\be\label{Quad}
\mathcal {H} = \sqrt{p^2 + 2(\alpha x + \beta y + \gamma z) + \alpha\beta-\gamma^2}
\ee
defines a three-parameter class of zero-birefringence relativistic NLEDs, where the parameters $(\alpha,\beta,\gamma)$ 
have dimensions of energy density. However, many distinct choices of these parameters yield theories that have equivalent 
field equations, and once this is taken into account we arrive at a list of six distinct possibilities.  In the nomenclature 
of \cite{Russo:2022qvz}, and including status with respect to electromagnetic duality  and convexity of the Hamiltonian density as a function of ${\bf D}$,  these six cases (all of which have one free parameter with dimensions of energy density, except the last which has no free parameters) are:

\begin{enumerate}

\item Born-Infeld (BI). $O(2)$ duality invariant. Strictly convex.

\item Pleba{\'n}ski (Pl). Not duality invariant. Not convex.

\item  reverse-BI (rBI) Not duality invariant. Not convex.

\item extreme-BI  (eBI). Not duality invariant. Convex.

\item magnetic-eBI (meBI). Not duality invariant. Convex,

\item  Bialynicki-Birula (BB). $Sl(2;\bR)$ duality invariant. Convex.

\end{enumerate}
Remarkably, this list is precisely the list of  {\sl all} known, or previously suspected, NLEDs 
with zero birefringence!  The appearance of eBI and meBI on this list confirms their zero birefringence status, and thus settles
the issue raised by the Lagrangian analysis in \cite{Russo:2022qvz}. As BB is duality invariant, its status in this respect was settled in \cite{Bandos:2023yat}.

Another purpose of this paper is to further investigate the special properties of the limits of BI, in particular the new ``extreme'' limits introduced in \cite{Russo:2022qvz}. One special feature already noted above, is the necessity for Lagrangian constraints; the reader may be wondering how this is compatible with equivalence of the Lagrangian and Hamiltonian formulations. Here it should be appreciated that the condition guaranteeing equivalence is 
convexity, of $\mathcal{L}$ as a function of ${\bf E}$ and of $\mathcal{H}$ as a function of ${\bf D}$. For example, given $\mathcal{H}$ the Lagrangian density is defined by the Legendre transform 
\be\label{legL}
\mathcal{L}({\bf E}, {\bf B}) := \sup_{{\bf D}} \left[{\bf D} \cdot {\bf E} 
- \mathcal{H}({\bf D},{\bf B})\right] \, ,    
\ee
which implies convexity of $\mathcal{L}$, and a further Legendre transform yields 
\be\label{legH}
\mathcal{H}({\bf D}, {\bf B}) := \sup_{{\bf E}} \left[ {\bf E} \cdot {\bf D} - \mathcal{L}({\bf E},{\bf B})\right] \, ,  
\ee
but this implies convexity of $\mathcal{H}$, which is therefore required for the Legendre transform to be involutive, and the same argument applies to $\mathcal{L}$. 

If $\mathcal{H}$ is not a convex function of ${\bf D}$ it is still  possible
to find a (formal) Lagrangian density (with Euler-Lagrange equations that are equivalent to the Hamiltonian field equations)
if the independent ${\bf D}$-field of the phase-space Lagrangian density $\widetilde{\mathcal{L}}$ of \eqref{start} is 
an auxiliary field that can be eliminated by using its field equation. This is the case for both Pleba{\'n}ski and 
reverse-BI but the non-convexity of  a NLED Hamiltonian density implies the existence of superluminal plane-wave disturbances for some constant uniform electromagnetic background \cite{Bandos:2021rqy,Russo:2022qvz}. For this reason, the Pleba{\'n}ski and rBI cases are unphysical. The remaining four cases, which are BI and its limits to eBI, meBI and BB, are all physical in this respect.

Although BI and its three limits all have a convex Hamiltonian density, only for BI is it ``strictly convex'', and this distinction is crucial to understanding why a Hamiltonian analysis yields more examples of zero-birefringence NLEDS than the (`standard') Lagrangian analyses. 
A sufficiently differentiable function is convex if its Hessian matrix is positive; i.e. has no negative eigenvalues. If all 
eigenvalues are positive the function is ``strictly convex''; this is the case for $\mathcal{H}_{\rm BI}$ and it implies that the 
equation ${\bf E} = \partial \mathcal{H}/\partial {\bf D}$ has a unique solution for ${\bf D}$ as a function of ${\bf E}$, such that
\eqref{legL} yields $\mathcal{L}_{\rm BI}$.  In effect, ${\bf D}$ is an auxiliary field in the `phase-space' Lagrangian density of \eqref{start}, as it is for the Pleba{\'n}ski and rBI cases  but $\mathcal{L}_{\rm BI}$ is the 
value of $\widetilde{\mathcal{L}}$ at a {\sl global maximum} (with respect to variation of ${\bf D}$) rather than a saddle point. 

The limits of BI all have the property that one or more of the eigenvalues of the Hessian matrix of $\mathcal{H}$ is zero, 
so that $\mathcal{H}$ is convex but not ``strictly convex''. As a consequence, the equation ${\bf E} = \partial \mathcal{H}/\partial {\bf D}$
no longer has a unique solution for ${\bf D}$.  In addition, this equation  imposes one or more constraints on ${\bf E}$;  one 
for each zero eigenvalue.  This is simply illustrated by the BB Hamiltonian density of \eqref{HamBB}; this is a 
convex function of ${\bf D}$ but its Hessian matrix has two zero eigenvalues; the corresponding Lagrangian constraints are
$S=0$ and $P=0$. If we impose these with Lagrange multipliers $(v,u)$ we get the `non-standard' BB Lagrangian density \cite{Bialynicki-Birula:1984daz}
\be
\mathcal{L}_{BB} = v S + u P\, . 
\ee
Conversely, by taking the two Lagrangian constraints into account in an application of \eqref{legH} one recovers the BB Hamiltonian density \cite{Bandos:2020hgy}. For the ``extreme'' limits of BI the Hamiltonian Hessian matrix has only one zero eigenvalue, and there is therefore only one Lagrangian constraint, which was found for both eBI and meBI in \cite{Russo:2022qvz}.   

In the following section we present our analysis of birefringence in the Hamiltonian formulation, focusing on relativistic NLED and the conditions for zero birefringence.  In section 3, we restrict to Hamiltonian densities for which $\mathcal{H}^2$ is a quadratic function of the rotation invariants $(x,y,z)$, showing that the joint conditions for Lorentz invariance and no birefringence restrict $\mathcal{H}$ to the form in \eqref{Quad}, and we show how this leads to the above list of six distinct zero-birefringence NLEDs, and why (extending the convexity/causality analysis of \cite{Russo:2022qvz} to generic stationary backgrounds) the only physical cases are BI and its three limits. 

In section 4 we present further details of the ``extreme'' BI limits introduced in \cite{Russo:2022qvz}, where wave propagation was analysed for static backgrounds.
We extend those results to generic stationary backgrounds, showing that the quadratic BI dispersion relation degenerates to a pair of linear dispersion relations, which become identical in the conformal BB limit. Another, but related, unusual feature of the ``extreme'' limits is that wave propagation is confined to a plane (which is reduced to a line for a static 
background). We show how this is explained by the unusual form of the extreme-BI stress-tensor. We also present a simplified derivation of the Lagrangian formulations of the electric/magnetic extreme-BI theories. 

It may not have escaped the reader's attention that we not yet mentioned the possibility  of additional zero-birefringence relativistic NLEDs for which $\mathcal{H}$ is {\sl not} of the special form \eqref{Quad}. We leave  discussion of this point, and some others, to a final summary section.

\section{Hamiltonian birefringence}\label{sec:Hambiref}

Recall that any pair of constant uniform electromagnetic fields $(\bar{\bf D},\bar{\bf B})$ solves the 
Hamiltonian field equations of \eqref{feqs} and \eqref{EHdefs}, and the homogeneous optical medium 
it provides is then a background in which any small-amplitude inhomogeneity will propagate. To  
expand the field equations about this background we write 
\be
 {\bf D} = \bar {\bf D} + {\bf d}\, , \qquad {\bf B} = \bar{\bf B} + {\bf b}\, .  
 \ee
 To first order in the perturbations $({\bf d},{\bf b})$, 
  \be
 {\bf E}= \bar {\bf E} + {\bf e} \, , \qquad {\bf H} = \bar{\bf H} + {\bf h}\, ,  
 \ee
 where 
 \be
 \begin{aligned} 
{\bf e}({\bf d},{\bf b}) &= \mathcal{H}_x {\bf d} + \mathcal{H}_z {\bf b} + 
\left[ \mathcal{H}_{xx} \bar{\bf D}\cdot{\bf d} +  \mathcal{H}_{xy}\bar{\bf B}\cdot{\bf b}
+ \mathcal{H}_{xz}\left(\bar{\bf D}\cdot{\bf b} + \bar{\bf B}\cdot{\bf d}\right) \right] \bar{\bf D} \\
&+ \left[ \mathcal{H}_{zx} \bar{\bf D}\cdot{\bf d} +  \mathcal{H}_{zy}\bar{\bf B}\cdot{\bf b}
+ \mathcal{H}_{zz}\left(\bar{\bf D}\cdot{\bf b} + \bar{\bf B}\cdot{\bf d}\right) \right]\bar{\bf B}\, , 
\end{aligned}
\ee
and 
\be
\begin{aligned}
{\bf h}({\bf d},{\bf b}) &= \mathcal{H}_y {\bf b} + \mathcal{H}_z {\bf d} + 
\left[ \mathcal{H}_{yx} \bar{\bf D}\cdot{\bf d} +  \mathcal{H}_{yy}\bar{\bf B}\cdot{\bf b}
+ \mathcal{H}_{yz}\left(\bar{\bf D}\cdot{\bf b} + \bar{\bf B}\cdot{\bf d}\right) \right] \bar{\bf B} \\
&+ \left[ \mathcal{H}_{zx} \bar{\bf D}\cdot{\bf d} +  \mathcal{H}_{zy}\bar{\bf B}\cdot{\bf b}
+ \mathcal{H}_{zz}\left(\bar{\bf D}\cdot{\bf b} + \bar{\bf B}\cdot{\bf d}\right) \right]\bar{\bf D}\, . 
\end{aligned}
\ee
The linearized field equations for $({\bf d},{\bf b})$ are 
 \be
 \begin{aligned}
 \dot {\bf d}  &= \boldsymbol{\nabla} \times {\bf h} \, , \qquad  \boldsymbol{\nabla} \cdot {\bf d} =0 \\
 \dot {\bf b} & = -\boldsymbol{\nabla}\times {\bf e} \, , \qquad \boldsymbol{\nabla} \cdot {\bf b} =0\, ,  
\end{aligned} 
\ee
where the constant coefficients involving derivatives of $\mathcal{H}$ are now functions of the constant uniform 
background fields.  For a plane wave with wave 4-vector $(\omega,{\bf k})$ the two dynamical field equations reduce to 
\be\label{wave-eqs}
\omega\, {\bf d}_0 +  {\bf k}\times {\bf h}_0 =0 \, , \qquad  \omega\, {\bf b}_0 - {\bf k}\times {\bf e}_0 =0\, ,  
\ee
where $({\bf d}_0, {\bf b}_0)$ are uniform constant 3-vector amplitudes, and 
\be\label{fromd}
{\bf e}_0 = {\bf e}({\bf d}_0,{\bf b}_0)\, , \qquad {\bf h}_0= {\bf h}({\bf d}_0,{\bf b}_0)\, .
\ee
The two constraint equations reduce to 
\be\label{fromc}
{\bf k} \cdot {\bf d}_0 =0\, , \qquad {\bf k} \cdot {\bf b}_0 =0\, ,  
\ee
but these are a consequence of \eqref{wave-eqs} unless $\omega=0$; their only effect is to eliminate two solutions of \eqref{fromd} with $\omega=0$. 
 
 We may rewrite \eqref{wave-eqs} in the form
 \be\label{Matrix}
\left(\begin{array}{cc} K^+ & L \\ -L'  & -K^- \end{array}\right)\left(\begin{array}{c} {\bf d}_0\\ {\bf b}_0\end{array}\right) =0\, , 
 \ee
 where $(K^\pm,L,L')$ are $3\times3$ matrices with the following entries
 \be
 \begin{aligned}
 K^+_{ij} =& \   \omega\, \delta_{ij} - \varepsilon_{ijk} k_k \mathcal{H}_z 
 + \mathcal{H}_{xz} ({\bf k}\times \bar{\bf D})_i \bar D_j + 
 \mathcal{H}_{yz} ({\bf k}\times \bar{\bf B})_i \bar B_j \\ 
 &+ \mathcal{H}_{xy} ({\bf k}\times \bar{\bf B})_i \bar D_j + \mathcal{H}_{zz} ({\bf k}\times \bar{\bf D})_i \bar B_j \\ \\
 K^-_{ij} =& -  \omega\, \delta_{ij} - \varepsilon_{ijk} k_k \mathcal{H}_z
 + \mathcal{H}_{xz} ({\bf k}\times \bar{\bf D})_i \bar D_j + 
 \mathcal{H}_{yz} ({\bf k}\times \bar{\bf B})_i \bar B_j \\ 
 &+ \mathcal{H}_{xy} ({\bf k}\times \bar{\bf D})_i \bar B_j + \mathcal{H}_{zz} ({\bf k}\times \bar{\bf B})_i \bar D_j \\ \\
  L_{ij} =& - \varepsilon_{ijk} k_k \mathcal{H}_y + \mathcal{H}_{yy} ({\bf k}\times \bar{\bf B})_i \bar B_j +
  \mathcal{H}_{zz} ({\bf k}\times \bar{\bf D})_i \bar D_j \\
  &+ \mathcal{H}_{yz} \left[ ({\bf k}\times \bar{\bf D})_i \bar B_j + ({\bf k}\times \bar{\bf B})_i \bar D_j\right] \\ \\
 L'_{ij} =& - \varepsilon_{ijk}k_k \mathcal{H}_x + \mathcal{H}_{xx} ({\bf k}\times \bar{\bf D})_i \bar D_j + 
 \mathcal{H}_{zz} ({\bf k}\times \bar{\bf B})_i \bar B_j \\
 &+ \mathcal{H}_{xz} \left[ ({\bf k}\times \bar{\bf D})_i \bar B_j + ({\bf k}\times \bar{\bf B})_i \bar D_j \right] \, . 
 \end{aligned}
 \ee
 A non-zero solution for the wave amplitudes requires 
\be\label{deterM}
\det M=0\, , \qquad M= \left(\begin{array}{cc} K^+ & L \\ -L'  & -K^- \end{array}\right)\, . 
\ee

 At this point it is useful to simplify the matrices $(K^\pm,L,L')$ by a suitable local choice of 3-space axes. 
 We may orient the axes such that 
 \be\label{axes}
 \begin{aligned}
 \bar {\bf D} =& \ (0,D_2,0)\, , \qquad \ \ D_2\ge0 \, , \\
 \bar {\bf B} = & \ (0,B_2,B_3)\, , \qquad B_3\ge0\, . 
 \end{aligned}
 \ee
  The background field momentum is then 
 \be
 \label{poynt}
\bar {\bf D} \times \bar{\bf B} = (p,0,0)\, , \qquad p=D_2B_3 \ge0\, . 
\ee
For this choice of axes,
\be
\begin{aligned}
K^+ &= \left(\begin{array}{ccc} \omega & -k_3 J + k_2 U & k_2 \hat H_z -k_3 V\\ k_3 \mathcal{H}_z & 
\omega- k_1 U &-k_1 \hat H_z \\
-k_2\mathcal{H}_z & k_1 J & \omega+ k_1 V \end{array}\right)  \\
K^- &= \left(\begin{array}{ccc} -\omega & -k_3 J + k_2 V & k_2 \hat H_z -k_3 U\\ 
k_3 \mathcal{H}_z & -\omega- k_1 V &-k_1 \hat H_z \\
-k_2\mathcal{H}_z & k_1 J & -\omega+ k_1 U \end{array}\right) \, , 
\end{aligned} 
\ee
and
\be
\begin{aligned}
L &= \left(\begin{array}{ccc} 0 & -k_3 R + k_2 Z & k_2 \hat H_y - k_3 Z \\ 
k_3\mathcal{H}_y & -k_1 Z & -k_1 \hat H_y \\
-k_2 \mathcal{H}_y & k_1 R & k_1 Z\end{array}\right)   \\
L'  &= \left(\begin{array}{ccc} 0 & -k_3 R' + k_2 Z' & k_2 \hat H_x - k_3 Z' \\ 
k_3\mathcal{H}_x & -k_1 Z' & -k_1 \hat H_x \\
-k_2 \mathcal{H}_x & k_1 R' & k_1 Z' \end{array}\right)\, , 
\end{aligned}
\ee
where 
\be\label{coeff1}
\begin{aligned}
J &= \mathcal{H}_z + D_2^2 \mathcal{H}_{xz} + B_2^2 \mathcal{H}_{yz} + B_2 D_2 \left(\mathcal{H}_{xy} 
+ \mathcal{H}_{zz}\right) \\
R &= \mathcal{H}_y + B_2^2 \mathcal{H}_{yy} + D_2^2 \mathcal{H}_{zz} + 2D_2 B_2 \mathcal{H}_{yz}\\
R' &= \mathcal{H}_x + B_2^2 \mathcal{H}_{zz} + D_2^2 \mathcal{H}_{xx} + 2D_2 B_2 \mathcal{H}_{xz} \\
\hat H_z &= \mathcal{H}_z + B_3^2 \mathcal{H}_{yz}\\
\hat H_x &= \mathcal{H}_x + B_3^2 \mathcal{H}_{zz} \\
\hat H_y &= \mathcal{H}_y + B_3^2 \mathcal{H}_{yy} \, , 
\end{aligned}
\ee
and 
\be\label{coeff2}
\begin{aligned}
U & = B_3\left(B_2 \mathcal{H}_{yz} + D_2 \mathcal{H}_{xy}\right)\\
V &=  B_3\left(B_2 \mathcal{H}_{yz} + D_2 \mathcal{H}_{zz}\right) \\
Z &= B_3 \left(B_2\mathcal{H}_{yy} + D_2 \mathcal{H}_{yz}\right) \\
Z' &= B_3\left(B_2\mathcal{H}_{zz} + D_2 \mathcal{H}_{xz}\right) \, . 
\end{aligned}
\ee

We now have the $6\times 6$ matrix $M$ expressed in terms of $\omega$ and several sets of coefficient functions for entries that include terms proportional to one or more components of the wave-vector ${\bf k}$. 
As we are principally interested in the determinant of $M$, it is convenient to permute its rows and columns 
to arrive at the following matrix {\sl with the same determinant}:
\be\label{Mprime}
M' = \left(\begin{array}{cccccc} \omega & 0 & -k_3J+ k_2 U & k_2 \hat H_z -k_3 V 
& -k_3R +k_2 Z & k_2 \hat H_y - k_3 Z \\ 
0 & \omega &  k_3 R' - k_2 Z' & -k_2 \hat H_x + k_3 Z' &  k_3 J - k_2 V & -k_2 \hat H_z + k_3 U\\
k_3 \mathcal{H}_z & k_3\mathcal{H}_y & \tilde\omega- k_1 X & -k_1 \hat H_z & -k_1 Z & -k_1 \hat H_y \\
-k_2\mathcal{H}_z & -k_2 \mathcal{H}_y &  k_1 J & \tilde\omega+ k_1 X & k_1 R & k_1 Z \\
-k_3\mathcal{H}_x & -k_3 \mathcal{H}_z &  k_1 Z' & k_1 \hat H_x  &  \tilde\omega+ k_1 X & k_1 \hat H_z \\
k_2 \mathcal{H}_x & k_2\mathcal{H}_z & -k_1 R' & -k_1 Z' & -k_1 J & \tilde\omega- k_1 X \end{array}\right)\, . 
\ee
We have used here the fact that 
\be
\omega -k_1 U =\tilde\omega -k_1 X\, , \qquad \omega +k_1 V =\tilde\omega +k_1 X\, , 
\ee
where
\be
X= \frac12(U+V) = B_3\left[B_2 \mathcal{H}_{yz}  + D_2(\mathcal{H}_{xy} + \mathcal{H}_{zz})\right]\, , 
\ee
and 
\be\label{def.omprime}
\tilde\omega = \omega - \frac12 k_1 (U-V) = \omega - \frac12 k_1p\left(\mathcal{H}_{xy}-\mathcal{H}_{zz}\right). 
\ee
Notice that $M'$ takes the form 
\be
M' = \left(\begin{array}{cc} \bb{A} & \bb{B} \\ \bb{C} & \bb{D} \end{array}\right)\, , 
\ee
where $\bb{A} = \omega \bb{I}_2$ and $\bb{D}$ is a $4\times 4$ matrix that is independent of both $k_2$ and 
$k_3$, while the rectangular matrices $\bb{C}$ and $\bb{B}$ are independent of both $\omega$ and $k_1$. 
In addition, the submatrices have the following properties:
\be
\begin{aligned}
(\omega,k_1) \to -  (\omega,k_1)   &\quad \Rightarrow \quad \bb{A} \to - \bb{A} \quad \& \quad \bb{D} \to - \bb{D}\, ,\\
(k_2,k_3) \to -(k_2,k_3) &\quad \Rightarrow \quad \bb{B}\to - \bb{B} \quad \&\quad \bb{C} \to -\bb{C}\, , 
\end{aligned} 
\ee
It follows from these properties that $\det M'$ is invariant under both $(\omega,k_1) \to -  (\omega,k_1)$ and  
$(k_2,k_3) \to -(k_2,k_3)$; this can be seen from either of the following identities:
\be\label{detids}
\left|\left| \begin{array}{cc} \bb{A} & \bb{B} \\ \bb{C} & \bb{D} \end{array}\right|\right| \equiv \left\{
\begin{array}{cc} (\det \bb{D})\det\left(\bb{A}- \bb{B}\bb{D}^{-1} \bb{C}\right)  & (\det \bb{D} \ne0) \\ 
(\det\bb{A}) \det\left(\bb{D} - \bb{C} \bb{A}^{-1} \bb{B} \right) & (\det \bb{A}\ne 0)  \end{array} \right. .
\ee

Although the matrix $M$ of \eqref{deterM} is a $6\times 6$ matrix only four of the six amplitudes 
$({\bf d}_0,{\bf b}_0)$ are physical because two are eliminated by the two constraints \eqref{fromc}, which 
apply only when $\omega=0$. This means that $\det M$, and hence $\det M'$,  must be a 6th-order polynomial in 
$\omega$ with two zero roots, and hence must take the form 
\be\label{detM'}
\det M' = \omega^2 P_4\, , 
\ee
where  $P_4$ is some quartic polynomial in $\omega$, or in $\tilde\omega$. If we choose to write $P_4$ as a polynomial in $\tilde\omega$ then the structure of $M'$ implies that the cubic term is missing; it can only come from 
the product of all six diagonal entries of $M'$ but their product yields a term of $P_4$ that is quadratic in 
$\tilde\omega^2$. Thus,  
\be
P_4 = \tilde\omega^4 - 2\Upsilon \tilde\omega^2 + 2\Xi \tilde\omega + \Omega\, , 
\ee
where $(\Upsilon,\Xi,\Omega)$ are expressions that are (quadratic, cubic, quartic) in the 
components of ${\bf k}$, subject to the above symmetries. In particular, $\Xi$ is odd under 
$k_1\to-k_1$ and therefore zero when $k_1=0$; it is also zero for static backgrounds
(for which $\tilde\omega=\omega$) because the residual rotational symmetry in the 
plane orthogonal to the background vector fields implies that $k_1$ appears only in the 
combination $(k_1^2+ k_3^2)$. 

For physical NLEDs we expect $P_4= P_2P_2'$ where $P_2$ and $P_2'$ are both real quadratic polynomials in $\tilde\omega$.  The dispersion relations for the two independent polarisations are 
then $P_2=0$ and $P_2'=0$. When $P_2\ne P_2'$  we have ``birefringence'', so the condition for zero
birefringence is $P_4=P_2^2$. When this occurs, $P_2=\tilde\omega^2 - \Upsilon$, which implies that 
$P_4(\tilde\omega)$ has no linear term. The zero-birefringence condition $P_4=P_2^2$ is therefore equivalent to the two conditions 
\be
\Xi=0\, , \qquad  \Upsilon^2 = \Omega \qquad ({\rm zero\ birefringence}).
\ee

It is instructive to see how the above conclusions can be verified by consideration of 
some special cases for which $P_4$ is easily found (we recall that  the condition \eqref{LI} 
may be used freely because of our restriction to relativistic theories):
\begin{itemize}

\item $k_1=0$. In this case 
$\bb{D}= \omega \bb{I}_4$. The first of the identities \eqref{detids} can now be used to show that 
$\det M'=\omega^2 P_4(\omega)$, where
\be
P_4 =  \det\left(\omega^2 \bb{I}_2 - \bb{B}\bb{C} \right)\, .
\ee
The $2\times 2$ matrix $\bb{B}\bb{C}$ is given by 
\be
\begin{aligned}
\bb{B}\bb{C} =& \quad  k_2^2 \left(\begin{array}{cc} (\hat H_y\mathcal{H}_x - \hat H_z\mathcal{H}_z) & 
(\hat H_y \mathcal{H}_z - \hat H_z \mathcal{H}_y) \\ (\hat H_x \mathcal{H}_z - \hat H_z \mathcal{H}_x)  & 
(\hat H_x\mathcal{H}_y - \hat H_z\mathcal{H}_z) \end{array}\right) \\
&+  k_3^2 \left(\begin{array}{cc} (R\mathcal{H}_x -J\mathcal{H}_z) & (R\mathcal{H}_z - J\mathcal{H}_y)\\
(R'\mathcal{H}_z -J \mathcal{H}_x) & (R'\mathcal{H}_y -J \mathcal{H}_z)  \end{array}\right) \\
&  -2k_2k_3 \left(\begin{array}{cc} (Z\mathcal{H}_x -X \mathcal{H}_z) & (Z\mathcal{H}_z -X\mathcal{H}_y) \\
(Z'\mathcal{H}_z -X\mathcal{H}_x) & (Z'\mathcal{H}_y - X \mathcal{H}_z) \end{array}\right)\, ,  
\end{aligned}
\ee
and this yields the result
\be\label{P4.1}
P_4 = \omega^4 - 2 \Upsilon \omega^2 + \Omega\, , 
\ee
where
\be
\begin{aligned}
\Upsilon =& \frac12 k_2^2 \left(\hat H_x \mathcal{H}_y + \hat H_y \mathcal{H}_x -2 \hat H_z\mathcal{H}_z\right) 
+ \frac12 k_3^3 \left( R\mathcal{H}_x + R'\mathcal{H}_y -2J\mathcal{H}_z\right) \\
& - k_2k_3 \left( Z\mathcal{H}_x + Z'\mathcal{H}_y -2X \mathcal{H}_z\right)\, , 
\end{aligned}
\ee
and 
\be
\begin{aligned}
\Omega =& k_2^4 \left(\hat H_x\hat H_y - \hat H_z^2\right) + k_3^4 \left(R R'-J^2\right) 
+ k_2^2 k_3^2 \left[ R\hat H_x + R'\hat H_y -2J\hat H_z + 4(ZZ'-X^2) \right] \\
& -2 k_2^3 k_3 \left( Z' \hat H_y + Z\hat H_x -2X \hat H_z\right) 
-2k_2k_3^3 \left(Z'R + ZR' -2XJ\right) \, . 
\end{aligned}
\ee
As expected,  $P_4$ is a quadratic function of $\omega$ because this is the only way that it can be 
invariant under $(\omega,k_1) \to -  (\omega,k_1)$ when $k_1=0$. 

\item $B_3=0$. In this case the background is static: $p=0$.  For $k_1=0$ the result of \eqref{P4.1} applies, 
but the residual rotation invariance now allows the $k_1$-dependence to be deduced from  the $k_3$-dependence. Thus, 
\be\label{P4static}
P_4 = \omega^4 - 2 \Upsilon \omega^2 + \Omega\, , 
\ee
but now with 
\be\label{staticUp}
\Upsilon = k_2^2  +  \frac12(R\mathcal{H}_x + R'\mathcal{H}_y - 2J\mathcal{H}_z)(k_1^2 + k_3^2) \, , 
\ee
and 
\be
\Omega =\Upsilon^2 - \frac14 \left[ (R\mathcal{H}_x + R'\mathcal{H}_y - 2J\mathcal{H}_z)^2 -
4(RR'-J^2)\right] (k_1^2+k_3^2)^2\, . 
\ee
In this static case, the condition for zero birefringence is 
\be\label{box}
\boxed{(R\mathcal{H}_x + R'\mathcal{H}_y - 2J\mathcal{H}_z)^2 = 4(RR'-J^2)}\, ,  
\ee
where the expressions for $(J,R,R')$ may now be written as 
\be\label{JRR'}
\begin{aligned}
J &= \mathcal{H}_z + 2x\mathcal{H}_{xz} + 2y \mathcal{H}_{yz} + z\left(\mathcal{H}_{xy} 
+ \mathcal{H}_{zz}\right) \, , \\
R &= \mathcal{H}_y + 2y\mathcal{H}_{yy} + 2x\mathcal{H}_{zz} + 2z \mathcal{H}_{yz}\, , \\
R' &= \mathcal{H}_x + 2y \mathcal{H}_{zz} + 2x \mathcal{H}_{xx} + 2z \mathcal{H}_{xz} \, .\\
\end{aligned}
\ee

This special case is sufficient for most choices of $\mathcal{H}$ that define 
a relativistic NLED because it is generically possible to Lorentz  boost to a frame in 
which the background is static. It is therefore a useful {\sl necessary} condition for the absence of birefringence
that we shall use in the following section.

\item $k_2=k_3=0$. In this case both rectangular matrices $\bb{B}$ and $\bb{C}$ are zero, so 
\be
\det M = \omega^2 \det \bb{D}\, , 
\ee
and hence  
\be
P_4 = \left|\left| \begin{array}{cccc} \tilde\omega- k_1 X & -k_1 \hat H_z & -k_1 Z & -k_1 \hat H_y \\
k_1 J & \tilde\omega+ k_1 X & k_1 R & k_1 Z \\
k_1 Z' & k_1 \hat H_x  &  \tilde\omega+ k_1 X & k_1 \hat H_z \\
 -k_1 R' & -k_1 Z' & -k_1 J & \tilde\omega- k_1 X \end{array}\right|\right|\, . 
\ee
A computation yields
\be\label{comp.P4}
P_4 =  \tilde\omega^4 - 2\tilde\omega^2 \Upsilon + 2\tilde\omega \Xi + \Omega\, , 
\ee
where 
\be
\begin{aligned} 
\Upsilon =& k_1^2 \left\{\frac12 \left(R\hat H_x + R' \hat H_y\right) -J\hat H_z - (ZZ'-X^2)\right\}\, , \\
\Xi =& k_1^3 \left\{ X[R\hat H_x - R'\hat H_y] - (RZ' -R'Z)\hat H_z - (Z\hat H_x -Z'\hat H_y)J\right\}\\
\Omega =&k_1^4 \left(\Delta_1 + \Delta _2\right)\, , 
\end{aligned}
\ee
with 
\be\label{Delta12}
\begin{aligned}
\Delta_1 = &\,   (ZZ'-X^2)^2 + (RR'-J^2)(\hat H_x\hat H_y - \hat H_z^2)  \\
& - (Z \hat H_x- X\hat H_z)(ZR' -XJ) - (Z' \hat H_y- X\hat H_z)(Z'R -XJ)\, , \\
\Delta_2 = &\, X\left[\hat H_x(JZ -XR) + \hat H_y (JZ'-XR')\right]  \\
& + \hat H_z\left[ X(RZ' + R'Z) -2J ZZ'\right]\, . 
\end{aligned}
\ee
The purpose of the decomposition of $\Omega$ in this last special case is 
that $\Delta_2=0$ for the special class of relativistic NLED that we consider in 
the following section. 

\end{itemize}

Because symmetries prevent the appearance of $k_1k_2$ and $k_1k_3$ terms in $\Upsilon$,  
we may deduce the generic expression for $\Upsilon$ by combining the results found above for the
 $k_1=0$ and $k_2=k_3=0$ special cases. This yields  
\be\label{genericUp}
\begin{aligned}
\Upsilon =& \frac12 k_1^2 \left\{R\hat H_x + R' \hat H_y -2J\hat H_z - 2(ZZ'-X^2)\right\}\\
&+ \frac12 k_2^2 \left(\hat H_x \mathcal{H}_y + \hat H_y \mathcal{H}_x -2 \hat H_z\mathcal{H}_z\right) 
+ \frac12 k_3^3 \left( R\mathcal{H}_x + R'\mathcal{H}_y -2J\mathcal{H}_z\right) \\
& - k_2k_3 \left( Z\mathcal{H}_x + Z'\mathcal{H}_y -2X \mathcal{H}_z\right)\, ,  
\end{aligned}
\ee
which reduces to $\Upsilon$ of \eqref{staticUp} in the case that $B_3=0$. Given a solution of the zero-birefringence conditions, this generic result for $\Upsilon$ determines the unique dispersion relation $P_2=0$ since, as noted above, $P_2=\tilde\omega^2 - \Upsilon$ whenever $P_4=P_2^2$.

\subsection{The stress-energy tensor}\label{subsec:stress}

In the preceding subsection, we have considered wave perturbations of generic 
homogeneous stationary backgrounds. The restriction to static backgrounds leads
to many simplifications, and we can expect almost all stationary backgrounds to be 
Lorentz boosts of some static background. However, exceptional cases can occur 
and we now turn to an analysis of which stationary backgrounds are {\sl not} the boost of any static background. At first sight, this appears to be a difficult problem to solve within a Hamiltonian framework because the Lorentz transformations of the fields $({\bf D},{\bf B})$ are nonlinear. However, there is a simple way around this problem via the stress-energy tensor.   

For any relativistic NLED with Hamiltonian density $\mathcal{H}({\bf D},{\bf B})$, the stress-energy tensor has components \cite{Bialynicki-Birula:1984daz}
\be
\Theta^{00} = \mathcal{H}\, , \qquad \Theta^{0i}= \Theta^{i0} = \left({\bf D} \times {\bf B}\right)^i\, , 
\ee
and 
\be
\Theta^{ij} = \delta^{ij}\left( {\bf E}\cdot {\bf D} + {\bf H}\cdot {\bf B} - \mathcal{H}\right)
- \left[E^i D^j + H^iB^j\right]\, ,  
\ee
which is symmetric because rotation invariance implies ${\bf E}\times {\bf D} + {\bf H} \times {\bf B} = {\bf 0}$. 
Equivalently, as a consequence of  \eqref{constit}, 
\be
\Theta^{ij} = \delta^{ij}(2W-\mathcal{H}) - \left[\mathcal{H}_x D^iD^i + 
\mathcal{H}_y B^iB^j + 2\mathcal{H}_z D^{(i}B^{j)}\right]\, , 
\label{tpm}
\ee
where
\be\label{W}
W := \frac12\left({\bf D}\cdot {\bf E} + {\bf B}\cdot {\bf H}\right) \equiv  x\mathcal{H}_x + y\mathcal{H}_y + z\mathcal{H}_z \, . 
\ee
At any given point in 3-space we may introduce an orthonormal basis $\{{\bf e}_i; i=1,2,3\}$ such that
\be
\begin{aligned}
{\bf p} =& \  p \, {\bf e}_1 \, , \quad (p\ge0)\\
{\bf D} =& \  D_2{\bf e}_2 \, , \quad (D_2\ge0)\\
{\bf B} =& \  B_2{\bf e}_2 + B_3{\bf e}_3\, , \quad (B_3\ge0)\, . 
\end{aligned}
\ee
The stress-energy tensor $\Theta^{\mu\nu}$ (for $\mu,\nu=0,1,2,3$) then takes block-diagonal form, and we may diagonalize the lower $2\times2$ block (locally) by means of an orthogonal transformation in the space spanned by  $({\bf e}_2,{\bf e}_3)$. We then have 
\be\label{stress-en}
\Theta = \left(\begin{array}{cccc} \mathcal{H} & p & 0 & 0 \\ p & 2W- \mathcal{H} & 0 & 0 \\ 0 & 0 & W_- -\mathcal{H}   & 0\\ 0 & 0 & 0 & W_+  - \mathcal{H} \end{array}\right) \, , 
\ee
where 
\be 
W_\pm = W \pm \sqrt{W^2-p^2}\, .
\ee
This expression is valid for any $\mathcal{H}$ that defines a Lorentz invariant theory; it follows that Lorentz invariance implies the inequality $W^2\ge p^2$.

If we take the trace of $\Theta$ with the Minkowski 
metric $\eta := {\rm diag.}(-1,1,1,1)$ we find that 
\be
\Theta^{\mu\nu} \eta_{\mu\nu} = 4(W-\mathcal{H}) \, , 
\ee
which is zero when $W=\mathcal{H}$; i.e. when $\mathcal{H}$ is a homogeneous degree-1 function of  the rotation invariants $(x,y,z)$. This is the condition for conformal invariance.  

In principle $W$ may have either sign but if $W<0$ then the pressure $(W_- -\mathcal{H})$ is negative with a magnitude greater than $\mathcal{H}$, which violates the Dominant Energy Condition. We may therefore expect a theory allowing $W<0$ to be unphysical. This is confirmed by the fact that convexity (of $\mathcal{H}$ as a function of ${\bf D}$)  implies $W\ge0$ (given Lorentz invariance) because convexity is required for causality;
we provide the details in subsection \ref{subsec:causality}. Given $W\ge0$, we may replace the inequality $W^2\ge p^2$ by the stronger inequality 
\be\label{ineqW}
W\ge p\, . 
\ee 

In general, \eqref{stress-en} is valid only at one chosen point in spacetime
but it is globally valid for uniform constant background fields, and in this case it is the stress-energy tensor for a homogeneous stationary optical medium, which is static when $p=0$. We wish to determine which non-static stationary backgrounds are Lorentz boosts of static backgrounds. An important fact about the stress-energy tensor is that it transforms linearly under Lorentz transformations, despite the nonlinear action of the Lorentz group on the fields $({\bf D},{\bf B})$.   
If $\Theta$ can be diagonalized by a Lorentz boost then the stationary background medium is a Lorentz boost of a static medium. As the lower $2\times2$ block of $\Theta$ is already diagonal, we may focus on the upper $2\times2$ block; call it $\Theta_{\rm up}$. In a boosted frame this becomes 
\be
\Theta_{\rm up}^\prime = L\Theta_{\rm up} L \,,  \qquad L = \left(\begin{array}{cc} \cosh\varphi & \sinh\varphi \\ \sinh\varphi & \cosh\varphi \end{array} \right)\, , 
\ee
where $\varphi$ is the boost parameter. This yields 
\be\label{boosted}
\Theta_{\rm up}^\prime = \left(\begin{array}{cc} \mathcal{H}'  & p' \\
p' & -\mathcal{H} + 2(\sinh\varphi\cosh\varphi)p + 2(\cosh^2\varphi) W \end{array}\right) \, , 
\ee 
where
\be
\begin{aligned}
\mathcal{H}'  &= \mathcal{H} + 2 (\sinh\varphi\cosh\varphi)p  + 2(\sinh^2\varphi) W\, , \\
p' &= (2\cosh^2\varphi -1) p + 2(\sinh\varphi\cosh\varphi)W\, . 
\end{aligned}
\ee
We need $\varphi$ such that $p'=0$, which requires 
\be\label{quadth}
p\tanh^2\varphi + 2W \tanh\varphi + p =0\, .  
\ee
This has the solution\footnote{Equivalently, $\tanh(2\varphi)=-p/W$.} 
\be\label{tanh}
\tanh\varphi = -\frac{W_-}{p}\,  , 
\ee
which has the property that $\tanh\varphi=0$ when $p=0$, as expected. The other solution does not have this property, and is singular at $p=0$, so we reject it. Using \eqref{tanh} in \eqref{boosted} we arrive at the following diagonal stress-energy tensor
\be\label{stress-en2}
\Theta^\prime = \left(\begin{array}{cccc} \mathcal{H} - W_-  & 0 & 0 & 0 \\ 0 & W_+ - \mathcal{H}   & 0 & 0 \\ 0 & 0 & W_- -\mathcal{H}   & 0\\ 0 & 0 & 0 & W_+ - \mathcal{H} \end{array}\right) \, , 
\ee
but this result assumes that \eqref{tanh} has a solution for {\sl finite} $\varphi$. This assumption is correct only if $W_-^2<p^2$, which is equivalent to
\be
(W_-)\sqrt{W^2-p^2} >0\, . 
\ee
This condition can fail to be satisfied in only two ways:  {\bf either} $W^2=p^2$ {\bf or}  $W_-=0$, but the latter option is possible only if $W^2=p^2$. Thus, the only backgrounds that are {\sl not} Lorentz boosts of some static background are those for which $W=p$, in which case 
\be\label{stationaryT}
\Theta = \left(\begin{array}{cccc} \mathcal{H} & p & 0 & 0 \\ p & 2p - \mathcal{H} & 0 & 0 \\ 0 & 0 & p -\mathcal{H}   & 0\\ 0 & 0 & 0 & p - \mathcal{H} \end{array}\right) \, .  \ee
As a check, we may return to \eqref{quadth} and set $W=p$ to deduce that 
$\tanh\varphi =-1$, which corresponds to an infinite boost; correspondingly, an infinite boost of the stress-energy tensor of \eqref{stationaryT} that takes $p\to0$ also takes $\mathcal{H}\to\infty$.

A case in which $W=p$ necessarily is $\mathcal{H}=p$; i.e. BB electrodynamics. As this is conformal, $W=\mathcal{H}=p$, and therefore $W=p$ for all (homogeneous) backgrounds, which are {\sl intrinsically} stationary. This is of course a very special case. In the following section, we investigate whether such backgrounds can occur in other models of interest that we now describe. 

\section{Relativistic `Quadratic' NLED}\label{sec:quad}

Any non-negative rotation invariant Hamiltonian density $\mathcal{H}$ can be written in the form $\mathcal{H}=\sqrt{f(x,y,z)}$ for some non-negative function $f$ of the rotation scalars $(x,y,z)$.
In terms of this function, the condition \eqref{LI} for Lorentz invariance is
\be\label{LIf}
f_xf_y-f_z^2 =4f\, .   
\ee
A particularly simple class of relativistic NLEDs can be found by choosing $f$ to be 
a quadratic polynomial; setting $(x,y,z)= (x^1,x^2,x^3)$ we now have
\be\label{quadf}
f= a_{ij} x^ix^j + b_i x^i + c\,   \qquad (i,j=1,2,3) 
\ee
for constant coefficients $(a,b,c)$. The constants $a_{ij}$ are dimensionless, while the $b_i$ 
have dimensions of energy density. Lorentz invariance imposes the following 
algebraic conditions on these coefficients: 
\be\label{alg1}
\begin{aligned}
a_{1 i} a_{2j} +a_{2 i} a_{1j} - 2a_{3i}a_{3j} &= 2a_{ij}\, , \\
b_1 a_{2i} + b_2 a_{1i} -2b_3a_{3i} &=2b_i\, ,\\
 b_1b_2-b_3^2 & =4c\, . 
\end{aligned}
\ee
Solutions of these equations yield what we shall call  (relativistic) ``quadratic'' NLEDs, but not all solutions
yield distinct theories.  The `geometric' term ${\bf E}\cdot {\bf D}$ in the phase-space Lagrangian density $\widetilde{\mathcal{L}}$ is unchanged by the field redefinition
\be
(A_0,{\bf A})\to  \rho (A_0,{\bf A})\, , \qquad {\bf D} \to \rho^{-1} {\bf D}\, , 
\ee
for any non-zero constant $\rho$; it is therefore a ``canonical'' transformation. 
In addition, ${\bf D}\to {\bf D} + \epsilon{\bf B}$ (for any constant $\epsilon$) adds a total derivative to $\widetilde{\mathcal{L}}$, so if we regard as equivalent any two phase-space actions that differ by a canonical transformation {\sl and} a possible total derivative,  then any two Hamiltonian densities $\mathcal{H}({\bf D},{\bf B})$ that 
are related by 
\be
{\bf D} \to \rho^{-1}{\bf D} + \epsilon {\bf B}\, , \qquad {\bf B} \to \rho {\bf B}\, , 
\ee
define equivalent NLEDs. 

Notice that it is consistent to set $b_i=0$ and $c=0$ in \eqref{alg1}.  The solutions for $a_{ij}$ then 
yield possible strong-field limits, which are conformal since the condition for conformal invariance (given Lorentz invariance) is ${\bf D}\cdot {\bf E} + {\bf B}\cdot {\bf H} = 2\mathcal{H}$  \cite{Bandos:2020jsw}, and this is equivalent to degree-2 homogeneity of $f(x,y,z)$. Once account is taken of equivalences, there are only four distinct 
possibilities, and the complete set of solutions to \eqref{alg1} may then be  organised according to which 
of these four strong-field limits applies:
\begin{enumerate}

\item $\boxed{f=(x+y)^2.}$ This yields the Maxwell Hamiltonian density: $\mathcal{H}_{\rm Max} =x+y$. 
The strong-field limit coincides with the weak-field limit. This remains true when we allow for non-zero $(b_i,c)$ 
because this leads to $f=(x+y +b)^2$, which just adds the constant $b$ to $\mathcal{H}_{\rm Max}$.

\item $\boxed{f= (x+y)^2 -z^2.}$ The corresponding Lagrangian density is 
\be
\mathcal{L} = \sqrt{S^2+P^2}\, , 
\ee
which is the interaction term of ModMax electrodynamics \cite{Bandos:2020jsw}. 
This is {\it not} a zero-birefringent theory because it does not correspond to a solution 
of the Boillat equations considered in \cite{Russo:2022qvz}. 

Allowing for non-zero $(b_i,c)$ we find in this case that \eqref{alg1} requires $b_1=b_2=2T$ for some constant $T$.
Setting $b_3=2gT$ for dimensionless constant $g$, we then find that
\be
f= (x+y+T)^2 - (z- gT)^2\, . 
\ee
The corresponding Lagrangian density is much more complicated. We will not need it here. 

\item $\boxed{f= 4xy.}$ In this case the Hessian matrix of $\mathcal{H}$ has one zero eigenvalue; the
corresponding Lagrangian constraint is $S=0$. As the `canonical' Lagrangian density is identically zero, 
the `non-standard' Lagrangian density obtained by imposing the constraint with a Lagrange multiplier $\lambda$ is
\be
\mathcal{L}= \lambda S\,  . 
\ee
Because the Lagrangian density is `non-standard' we cannot use the results of \cite{Russo:2022qvz}
to determine its birefringence properties. We return to this point below. 

Let us check the above result for $\mathcal{L}$ by using it to compute $\mathcal{H}$. We first observe that
\be
{\bf D} := \frac{\partial\mathcal{L}}{\partial{\bf E}} = \lambda {\bf E} \qquad (\Rightarrow \ {\bf E}= \lambda^{-1}{\bf D}) 
\ee
and hence 
\be
\mathcal{H}_\lambda := {\bf D}\cdot {\bf E} - \mathcal{L} = \lambda^{-1} x + \lambda y\, .  
\ee
Eliminating $\lambda$, which is now an auxiliary field, we find that $\mathcal{H}_\lambda \to \sqrt{4xy}$. 

Allowing for non-zero $(b_i,c)$ we find that 
\be\label{Case3gen}
f= (2x+T)(2y+T)\, . 
\ee
For $T\ge0$ the Hamiltonian is convex and the Hessian matrix has no zero eigenvalues if $T>0$. This tells us that there must be a standard Lagrangian; it is 
\be
\mathcal{L} = - \sqrt{T(T-2S)}\, , 
\ee
which is the first example of a nonlinear extension of Maxwell electrodynamics, introduced by Born in 1933 \cite{Born:1933qff}. This is {\sl not} a zero-birefringence NLED. We shall soon see that its conformal strong-field limit is also birefringent.

\item $\boxed{f= 4xy-z^2 \equiv p^2.}$ In this case $\mathcal{H}= |{\bf D}\times {\bf B}|$, which defines BB electrodynamics. 
Allowing for non-zero $(b_i,c)$ we find that \eqref{alg1} imposes no conditions on the $b_i$. Thus, after the relabelling
$(b_1,b_2,b_3) = 2(\alpha,\beta,\gamma)$, we find that 
\be\label{Qdef}
f= Q(x,y,z) := 4xy -z^2 + 2(\alpha x+ \beta y + \gamma z) + \alpha\beta-\gamma^2 \, . 
\ee
We shall discuss this case in detail in most of what follows. 

\end{enumerate}

Our results of the previous section can be used to determine the birefringence properties
of all the above ``quadratic'' relativistic NLEDs. However, we begin by focusing on the 
four strong-field limits for which $f$ is a {\sl homogeneous} quadratic in $(x,y,z)$. Because of the 
homogeneity of $f$, the coefficient functions $(J,R,R')$ defined in \eqref{JRR'} reduce to
\be
\begin{aligned}
J &=\  \frac{1}{2\mathcal{H}}\left[f_z + z(f_{xy}- f_{zz}-2)\right]\, , \\
R &= \  \frac{1}{2\mathcal{H}}\left[f_y -2x (f_{xy}- f_{zz}-2)\right]\, , \\
R' &= \  \frac{1}{2\mathcal{H}}\left[f_x -2y (f_{xy}- f_{zz}-2)\right]\, ,  
\end{aligned}
\ee
and this leads to the conclusion that the unique solution of \eqref{box} is $f= 4xy-z^2$. 

This settles the birefringence status of the third case above ($f=4xy$): it is birefringent, and 
hence so are the related non-conformal cases with $f$ given by \eqref{Case3gen}. This is
also true for the second case, as already noted.  It does {\sl not}  settle the birefringence status 
of the fourth case ($f=p^2$) for two reasons. One is that, since $W=\mathcal{H}=p$, all backgrounds with $\mathcal{H}\ne0$ are stationary and not static, but \eqref{Case3gen} applies only for static backgrounds. The other is that birefringence  could disappear in the strong-field limit.  

However, we {\sl can} conclude from this test based on \eqref{Case3gen} that all {\sl interacting}  ``quadratic'' relativistic NLEDs without birefringence have a Hamiltonian density of the form $\mathcal{H}=\sqrt{Q}$, where $Q$ is the three-parameter family of functions defined in \eqref{Qdef}. For most of the remainder of this section, we shall apply the general birefringence results of section \ref{sec:Hambiref} to the NLED class defined by this Hamiltonian density. 

\subsection{Birefringence redux} \label{biref-redux}

A useful equivalent form of $\mathcal{H}=\sqrt{Q}$ is 
\be\label{equivH}
\mathcal{H} = \sqrt{(2x+\beta)(2y+\alpha) - (z-\gamma)^2}\, . 
\ee
The first derivatives are 
\be\label{firstd}
\mathcal{H}_x = \frac{(2y+\alpha)}{\mathcal{H}}\, , \qquad \mathcal{H}_y = \frac{(2x+\beta)}{\mathcal{H}}\, , \qquad 
\mathcal{H}_z = -\frac{(z-\gamma)}{\mathcal{H}}\, ,  
\ee
and the second derivatives are 
\bea
\mathcal{H}_{xx} &=& - \frac{(2y+\alpha)^2}{\mathcal{H}^3}\,, \qquad \qquad\
\mathcal{H}_{yy} = - \frac{(2x+\beta)^2}{\mathcal{H}^3}\,, \nonumber \\
\mathcal{H}_{xz} &=& \frac{(2y+\alpha)(z-\gamma)}{\mathcal{H}^3} \, , \qquad \ \
\mathcal{H}_{yz} = \frac{(2x+\beta)(z-\gamma)}{\mathcal{H}^3} \, , \\
\mathcal{H}_{xy} &=&  \frac{\mathcal{H}^2 -(z-\gamma)^2}{\mathcal{H}^3}\,,  \qquad \quad
\mathcal{H}_{zz} = - \frac{\mathcal{H}^2 +(z-\gamma)^2}{\mathcal{H}^3}\,. \nonumber
\eea

Using these formulae, we may compute expressions for the various coefficient functions
defined in \eqref{coeff1} and \eqref{coeff2}:
\be
\frac{(U-V)}{2}  =  \frac{p}{\mathcal{H}}\, , 
\ee
and
\bea
 X &=& C \mathcal{H}_z \, , \qquad \ \ Z=  C\mathcal{H}_y \, , \quad \quad \
 Z' = C \mathcal{H}_x \, , \nonumber  \\
J &=& A \mathcal{H}_z \, , \qquad \ \  R= A\mathcal{H}_y \, , \quad \quad \ R'= A\mathcal{H}_x\, ,  \\
\hat H_x &=& K\mathcal{H}_x\, , \qquad \hat H_y= K\mathcal{H}_y\, , 
\qquad \hat H_z= K\mathcal{H}_z \nonumber \, , 
\eea
where the constants $(A,K,C)$ are 
\smallskip
\be
A= \frac{(\alpha\beta-\gamma^2) + \beta B_3^2}{\mathcal{H}^2}\, , \qquad 
K = 1- \frac{[p^2 + \beta B_3^2]}{\mathcal{H}^2} \, , \qquad  
C =  -\frac{\left(\gamma p + \beta B_2B_3\right)}{\mathcal{H}^2}\, .  
\ee
\smallskip

Using these results in the matrix $M'$ of \eqref{Mprime} we have
\be\label{Mprime'}
M' = \left(\begin{array}{cccccc} \omega & 0 & \kappa \mathcal{H}_z +\frac{k_2 p}{\mathcal{H}}  & 
\tau\mathcal{H}_z +\frac{k_3 p}{\mathcal{H}}   & 
\kappa \mathcal{H}_y & \tau\mathcal{H}_y  \\ 
0 & \omega &  -\kappa\mathcal{H}_x & -\tau\mathcal{H}_x & -\kappa\mathcal{H}_z + \frac{k_2 p}{\mathcal{H}} & 
-\tau\mathcal{H}_z + \frac{k_3 p}{\mathcal{H}} \\
k_3 \mathcal{H}_z & k_3\mathcal{H}_y & \tilde\omega_-  & -k_1K \mathcal{H}_z & -k_1 C\mathcal{H}_y & -k_1 K\mathcal{H}_y \\
-k_2\mathcal{H}_z & -k_2 \mathcal{H}_y &  k_1 A\mathcal{H}_z & \tilde\omega_+ & k_1 A\mathcal{H}_y & k_1 C\mathcal{H}_y \\
-k_3\mathcal{H}_x & -k_3 \mathcal{H}_z &  k_1 C \mathcal{H}_x & k_1 K\mathcal{H}_x  &  
\tilde\omega_+  & k_1 K \mathcal{H}_z \\
k_2 \mathcal{H}_x & k_2\mathcal{H}_z & -k_1A\mathcal{H}_x & -k_1 C\mathcal{H}_x & -k_1 A\mathcal{H}_z  & \tilde\omega_- \end{array}\right) 
\ee
where 
\be
\kappa =Ck_2 -Ak_3 \, , \qquad  \tau = Kk_2 -C k_3 \, , 
\ee 
and 
\be
\tilde\omega_\pm  = \tilde\omega \pm k_1 C\mathcal{H}_z\, , \qquad \tilde\omega = \omega - \frac{k_1p}{\mathcal{H}} \, .
\ee
It is now straightforward to show that $P_4=P_2^2$ for all  of the special cases (of wave-vector and background) considered in the previous section:
\begin{itemize}

\item $k_1=0$. In this case\footnote{Care must be taken in using this result in the context of small-amplitude wave propagation because the full dispersion relation has a term linear in $k_1$ (unless $p=0$) that contributes to the group velocity even when $k_1=0$.}
\be
\Upsilon = K k_2^2 + A k_3^2 -2C k_2k_3 \, , \qquad \Omega = \Upsilon^2\, . 
\ee 
It follows that $P_4=P_2^2$.

\item $B_3=0$. In this (static) case $p=0$, and
\be
C=0\, , \qquad A= A_0\equiv \frac{\alpha\beta -\gamma^2}{\mathcal{H}^2}\, , \qquad K= 1\, , 
\ee
and this yields
\be\label{staticUp2}
\Upsilon = k_2^2 + A_0(k_1^2+ k_3^2) \, , \qquad \Omega= \Upsilon^2\, . 
\ee 
It again follows that $P_4=P_2^2$. 

\item $k_2=k_3=0$. In this case 
\be
\Upsilon = (AK-C^2) k_1^2 \, , \qquad \Xi=0\, , \qquad \Omega=\Upsilon^2\, , 
\ee
where the result for $\Omega$ is a consequence of 
\be
\Delta_1 = (AK-C^2)^2\, , \qquad \Delta_2=0\, . 
\ee
Again, $P_4=P_2^2$.

\end{itemize}

These results suggest that $P_4=P_2^2$ for all relativistic NLEDS with $\mathcal{H}=\sqrt{Q}$, {\sl irrespective of the choice of background} (or wave-vector ${\bf k}$). This must be true for those parameter choices for which 
all stationary backgrounds are Lorentz boosts of some static background, but 
we wish to allow for the possibility of intrinsically-stationary backgrounds
that are not the Lorentz boost of any static background. This is because of the 
{\it a priori} possibility of NLEDs that exhibit birefringence {\sl only} in such backgrounds. 

We observed earlier that all backgrounds of BB electrodynamics are intrinsically stationary, and we shall see later that some BI backgrounds are also of this type. In these two cases we know from earlier work that $P_4=P_2^2$ even in these cases \cite{Bialynicki-Birula:1984daz,Bandos:2023yat}, and we shall see later that the ``extreme'' limits of BI do not allow intrinsically-stationary backgrounds. For present purposes, therefore, there is no need for a detailed analysis of the many zero-birefringence conditions for the stationary case. Using Mathematica, we have verified that all of them are solved by $\mathcal{H}=\sqrt{Q}$ for all values of the parameters $(\alpha, \beta,\gamma)$. However, we give here the explicit form of one of these conditions because of its simple structure:
\be\label{detcon}
\left|\left| \begin{array}{ccc} \mathcal{H}_z & \mathcal{H}_y & \mathcal{H}_x \\
\mathcal{H}_{xz} & \mathcal{H}_{zz} & \mathcal{H}_{xx} \\
\mathcal{H}_{yz} & \mathcal{H}_{yy} & \mathcal{H}_{zz} 
\end{array}\right|\right| =0\, . 
\ee
This is very different from the condition \eqref{box}, at least superficially, but it is easily verified that it is solved by $\mathcal{H}=\sqrt{Q}$.  Its simplicity and form suggest an underlying geometrical interpretation that might make possible an explicit solution of the Hamiltonian zero-birefringence equations, as is possible if electromagnetic duality invariance is assumed \cite{Bandos:2023yat}. 

To summarise: {\sl any Hamiltonian density of the form \eqref{Quad} defines a zero-birefringence NLED}. The unique dispersion relation is 
\be\label{UpQ}
\tilde\omega^2 = (AK-C^2)k_1^2 + \Phi(k_2,k_3)\, , \qquad 
\tilde\omega = \omega - \frac{k_1p}{\mathcal{H}}\, ,  
\ee
where $\Phi$ is the quadratic form 
\be\label{Phi-form}
\Phi := Kk_2^2 + Ak_3^2 -2C k_2 k_3 \, .
\ee
This is the specialisation to $\mathcal{H}=\sqrt{Q}$ of  \eqref{genericUp}, and \eqref{def.omprime}. 

For future use we note here that $\Phi\ge0$. This follows from the fact that the associated 
$2\times2$ matrix has trace $(A+K)$ and determinant $(AK-C^2)$, and both are positive since 
\be\label{CKA-form}
AK -C^2= A_0 \, , \qquad A+K= 1 - \frac{p^2}{\mathcal{H}^2} + A_0\, . 
\ee
where $A_0$ is $A$ for a static background:
\be 
\label{staticA0}
A_0= \frac{(\alpha\beta -\gamma^2)}{\mathcal{H}^2}\, .
\ee

Using the first of the relations \eqref{CKA-form}, we may rewrite the dispersion relation of \eqref{UpQ} in the form
\be\label{Unique}
\boxed{\tilde\omega^2 = A_0 k_1^2 + \Phi(k_2,k_3)} \, . 
\ee
This equation will be the starting point for our analysis of properties of small-amplitude wave propagation for the $\mathcal{H}=\sqrt{Q}$ class of NLEDs. For Born-Infeld, for example, it is the equation
\be\label{BI-wave}
\left(\omega - \frac{{\bf k}\cdot{\bf p}}{\mathcal{H}_{\rm BI}}\right)^2 = \frac{1}{\mathcal{H}_{\rm BI}^2}\left\{T^2|{\bf k}|^2 
+ T\left[({\bf k}\cdot {\bf D})^2 + ({\bf k}\cdot {\bf B})^2\right]\right\}\, .
\ee
For a static background, \eqref{Unique} reduces to 
\be\label{static-disp}
\omega^2 = k_2^2 + A_0(k_1^2+k_3^2)  \qquad (p=0). 
\ee

\subsection{Inequivalent zero-birefringence NLEDs}\label{sec:Inequiv}

We have now seen that all members of the three-parameter class of Hamiltonian densities $\mathcal{H}=\sqrt{Q}$ 
define a zero-birefringent NLED. In terms of the gauge-invariant 
Hamiltonian fields $({\bf D},{\bf B})$, 
\be\label{quadclass}
\mathcal{H}= \sqrt{\left|{\bf D}\times {\bf B}\right|^2 + \left(\alpha |{\bf D}|^2 + \beta |{\bf B}|^2 + 2\gamma {\bf D}\cdot {\bf B}\right)
+ (\alpha\beta -\gamma^2)}\,  .
\ee
However not all choices of the parameters $(\alpha,\beta,\gamma)$ yield physically distinct theories, for reasons already explained. To investigate this we shall need to separate those cases for which $\alpha$ is non-zero from those for which $\alpha$ is zero: 

\begin{itemize}

    \item $\boxed{\alpha\ne0.}$ In this case we may redefine ${\bf D}$ by a shift
    \be\label{Dshift}
    {\bf D} \to {\bf D} - \epsilon {\bf B}\, . 
    \ee 
    This shifts the  ${\bf E}\cdot {\bf D}$ term in the phase-space Lagrangian by a multiple of ${\bf E}\cdot {\bf B}$, 
    but this is a total derivative that we ignore. The shift of ${\bf D}$ also changes $Q$ but not the leading 
    $\left|{\bf D}\times {\bf B}\right|^2$ term. For the choice $\epsilon = \gamma/\alpha$ we find that 
    \be 
    Q \ \to \ \left|{\bf D}\times {\bf B}\right|^2 + \left[\alpha |{\bf D}|^2 + \frac{(\alpha\beta-\gamma^2)}{\alpha}\, |{\bf B}|^2\right]
    + (\alpha\beta - \gamma^2)\, . 
    \ee
We now consider separately the three subcases for which the constant term in this expression is
positive, zero or negative:

    \begin{enumerate}

    \item $\alpha\beta-\gamma^2>0$. In this subcase we perform the following 
    rescaling:
\be\label{rescaling}
({\bf D},{\bf B}) \ \to\  \left(\lambda^{-1} {\bf D}, \lambda{\bf B}\right)\, ,
\ee
where the rescaling of ${\bf B}$ is induced by a rescaling ${\bf A}\to \lambda{\bf A}$, which 
leaves ${\bf E}\cdot {\bf D}$ unchanged if we similarly rescale $A_0\to \lambda A_0$. 
If the scaling parameter $\lambda$ is chosen such that
\be
\lambda^2 = \frac{|\alpha|}{\sqrt{\alpha\beta-\gamma^2}}\, , 
\ee
then $Q\to Q_{\rm BI}$, where 
\be
Q_{\rm BI} = \left|{\bf D}\times {\bf B}\right|^2 + T(|{\bf D}|^2 + |{\bf B}|^2) + T^2 \, , \qquad \left(T= \pm\sqrt{\alpha\beta-\gamma^2}\right). 
\ee
For $T>0$ this yields the BI Hamiltonian. \\

\item $\alpha\beta-\gamma^2=0$. In this subcase $Q\to Q_{\rm eBI}$, where
\be
Q_{\rm eBI} = \left|{\bf D}\times {\bf B}\right|^2 + T|{\bf D}|^2 \, , \qquad \left(T=\alpha \right)
\ee
This yields, for $T>0$, the Hamiltonian of the (electric) ``extreme'' limit of  Born-Infeld (eBI). \\

\item $\alpha\beta-\gamma^2 <0$. In this subcase we may perform the rescaling of
\eqref{rescaling} but now with 
\be
\lambda^2 = \frac{|\alpha|}{\sqrt{\gamma^2-\alpha\beta}}\, .   
\ee
Now $Q\to Q_{\rm rBI}$, where
\be\label{newRBI}
Q_{\rm rBI} = \left|{\bf D}\times {\bf B}\right|^2+ 2T (|{\bf D}|^2 -|{\bf B}|^2) - T^2 \, ,\qquad 
\left(T= \pm\sqrt{\gamma^2-\alpha\beta}\right). 
\ee
This yields (one form of) the Hamiltonian density for ``reverse Born-Infeld'' (rBI). Notice that positivity of $Q$, required for reality of $\mathcal{H}$, imposes some lower limit on $|{\bf D}|-|{\bf B}|$, assuming $T>0$. If $T<0$ then the lower limit is on 
$|{\bf B}|-|{\bf D}|$ (so there is an electric and magnetic version of rBI, according to the choice of sign for $T$). 

The rBI Hamiltonian density was given in \cite{Russo:2022qvz} in an alternative but equivalent form. 
To see the equivalence we return to \eqref{quadclass} and perform the ${\bf D}$-shift of \eqref{Dshift} but now choosing 
$\epsilon$ such that
\be
\alpha\epsilon= \gamma \pm \sqrt{\gamma^2-\alpha\beta}\, . 
\ee
This eliminates the term in $Q$ that is linear in $y$, and we now get
\be
Q'_{\rm rBI} = \left|{\bf D}\times {\bf B}\right|^2 + T|{\bf D}|^2 + 2\kappa {\bf D}\cdot{\bf B} - \kappa^2\, , 
\ee
where
\be
T=\alpha\, , \qquad \kappa = \pm \sqrt{\gamma^2- \alpha\beta}\, . 
\ee
This yields the rBI Hamiltonian density in the form given in \cite{Russo:2022qvz}; this form allows an obvious limit to Pleba{\'n}ski electrodynamics, which is one of the $\alpha=0$ subcases to which we now turn our attention. \\

\end{enumerate}

\item $\boxed{\alpha =0.}$ We again have three subcases to consider:

\begin{enumerate}

\item $\gamma\ne0$. In this subcase we may redefine ${\bf D}$ by the shift
\be
{\bf D} \to {\bf D} - \frac{\beta}{2\gamma} \, {\bf B}\, ,  
\ee
which results in $Q\to Q_{\rm Pl}$, where
\be
Q_{\rm Pl} = \left|{\bf D}\times {\bf B}\right|^2 + 2\kappa {\bf D}\cdot{\bf B} - \kappa^2 \, , \qquad (\kappa = \gamma). 
\ee
This yields the Hamiltonian of ``Pleba{\'n}ski'' electrodynamics found in \cite{Russo:2022qvz}, by Legendre transform of 
Pleba{\'n}ski's  Lagrangian density $\mathcal{L}_{\rm Pl} \propto S/P$ \cite{Plebanski:1970zz}.
Notice that positivity of $Q$ (required for reality of $\mathcal{H}$) imposes a restriction on the domain of the function $Q$. Assuming (without loss of generality) that $\kappa>0$, this restriction is  
\be
\frac{z}{\sqrt{4xy}} \ge -1 + \frac{\kappa}{\sqrt{4xy}}\, .
\ee
\\

\item $\gamma=0$. In this subcase we have
\be
Q = Q_{\rm meBI} = \left|{\bf D}\times {\bf B}\right|^2 + T|{\bf B}|^2 \, , \qquad (T= \beta)\, . 
\ee
For $T>0$ this yields the Hamiltonian density proposed in \cite{Russo:2022qvz} as a `magnetic'  extreme limit of Born Infeld (meBI). \\

\item $\gamma=0$ and $\beta=0$. In this case 
\be
Q= Q_{\rm BB} = \left|{\bf D}\times {\bf B}\right|^2 \, . 
\ee
This yields the Hamiltonian density of the conformal, but interacting, electrodynamics of 
Bialynicki-Birula \cite{Bialynicki-Birula:1984daz}.

\end{enumerate} 

\end{itemize}
We have now recovered the classification of zero-birefringence NLEDs found
from the Lagrangian approach to this problem, and we have extended it with 
proofs of the zero-birefringence status of the three limits of BI, in particular the
electric and magnetic ``extreme'' limits for which this status was not previously 
settled. 

The next step is to separate those zero birefringence NLEDs that are 
physical from those that are unphysical.

\subsection{Convexity and Causality}\label{subsec:causality}

 The importance of convexity (of the Lagrangian density as a function of ${\bf E}$ and the Hamiltonian density as a function of ${\bf D}$) was summarized in the Introduction. It guarantees the equivalence of the Lagrangian and Hamiltonian formulations. It is also required to eliminate the possibility of superluminal propagation.
  
 Convexity of the Hamiltonian density is equivalent to positivity of its $3\times3$ Hessian matrix, which has entries
 \be
 \bb{H}_{ij} := \frac{\partial^2\mathcal{H}}{\partial D^i \partial D^j} \ =\ 
 \mathcal{H}_x \delta_{ij} + \mathcal{H}_{xx} D_iD_j + 2\mathcal{H}_{xz} D_{(i} B_{j)} + \mathcal{H}_{zz} B_iB_j\, . 
 \ee
 As $\bb{H}$ is a symmetric matrix, all three eigenvalues are real, and if none of them is negative then $\bb{H}$ is positive; zero eigenvalues are permissible but if all are positive then $\bb{H}$ is ``strictly positive''. The three eigenvalues of 
 $\bb{H}$ are
 \be
h_0= \mathcal{H}_{x}\ ,\quad h_\pm =\mathcal{H}_{x}+ \Lambda  \ \pm \sqrt{\Lambda^2  - 
\left(\mathcal{H}_{xx}\mathcal{H}_{zz} - \mathcal{H}_{xz}^2\right) p^2}\, , 
\ee
where $p^2 =4xy-z^2$, and 
\be
\Lambda = x\mathcal{H}_{xx} + y \mathcal{H}_{zz} + z\mathcal{H}_{xz}\, . 
\ee
It follows that convexity requires both $\mathcal{H}_x\ge0$ {\sl and} 
\be\label{convex-ineq}
\mathcal{H}_x^2 + 2\Lambda \mathcal{H}_x + 
\left(\mathcal{H}_{xx}\mathcal{H}_{zz} - \mathcal{H}_{xz}^2\right) p^2 \ge0\, . 
\ee
These results apply for any Hamiltonian density, but if we insist on 
Lorentz invariance then we also require \eqref{LI}, which can only be satisfied if $\mathcal{H}_x\mathcal{H}_y>0$. Thus convexity combined with Lorentz invariance requires, in addition to \eqref{convex-ineq}, 
\be\label{C+L}
\mathcal{H}_x >0 \, , \qquad \mathcal{H}_y >0\, . 
\ee

We claimed in subsection \ref{subsec:stress} that convexity and Lorentz invariance combined imply that the function $W$ defined in \eqref{W} is non-negative. The proof is as follows. The inequalities \eqref{C+L} allow us to rewrite the equation $W\ge0$ in the form 
\be
(x\mathcal{H}_x+y\mathcal{H}_y)^2 \ge (z\mathcal{H}_z)^2\, . 
\ee
By adding $-4xy\mathcal{H}_x\mathcal{H}_y$ to both sides, and using \eqref{LI} to eliminate $\mathcal{H}_z$, we may rewrite this inequality as 
\be
(x\mathcal{H}_x-y\mathcal{H}_y)^2 +z^2 + p^2\mathcal{H}_x\mathcal{H}_y \ge0 \, ,   
\ee
which is satisfied, because all terms on the left-hand side are non-negative, and saturated only in the vacuum. 

Now we turn to the special cases with $\mathcal{H}=\sqrt{Q}$. This choice yields 
\be
\Lambda = - \frac{\mathcal{H}_x}{\mathcal{H}^2} \left[ p^2 +\alpha x+ \beta y + \gamma z\right] \, , \qquad 
\left(\mathcal{H}_{xx}\mathcal{H}_{zz} - \mathcal{H}_{xz}^2\right)=  \frac{\mathcal{H}_x^2}{\mathcal{H}^2}\, ,  
\ee
and \eqref{convex-ineq} reduces to 
\be
(\alpha\beta-\gamma^2) \mathcal{H}_x^2 \ge 0\, . 
\ee
As $\mathcal{H}=\sqrt{Q}$ defines a Lorentz invariant NLED, 
\eqref{C+L} applies and the convexity conditions for this case are 
\be\label{convex2}
\alpha\ge 0 \,,  \quad \beta \ge 0, \qquad \alpha\beta-\gamma^2 \ge0 \, .  
\ee
For $\alpha\beta>0$ we have BI. For $\alpha\beta=0$ but $\gamma\ne0$ we have the extreme limits of BI. For $\alpha=\beta=\gamma=0$ we have BB. To summarize, only the following four of the six zero-birefringence NLEDs have a convex Hamiltonian density:
\begin{itemize}

\item BI.  Hessian of $\mathcal{H}$ has no zero eigenvalues. 

\item eBI.  Hessian of $\mathcal{H}$ has one zero eigenvalue

\item meBI. Hessian of $\mathcal{H}$ has one zero eigenvalue

\item BB.  Hessian of $\mathcal{H}$ has two zero eigenvalues.

\end{itemize}
For the other two zero-birefringence NLEDs (Pleba{\'n}ski and reverse-BI) the Hessian of  $\mathcal{H}$ has at least one negative eigenvalue, which allows superluminal propagation on some constant electromagnetic backgrounds \cite{Russo:2022qvz}. We now revisit this issue from a Hamiltonian perspective, and extend previous results to
include generic stationary backgrounds. 

Recall that, for $\mathcal{H}=\sqrt{Q}$, the dispersion relation for wave propagation in 
a static background is given by \eqref{static-disp}, and the wave group-velocity $v_g$ is therefore given by 
\be\label{vg}
v_g^2 = \left| \frac{d\omega}{d{\bf k}}\right|^2 = 
\frac{A_0^2|{\bf k}_\perp|^2 + 
k_\parallel^2}{A_0|{\bf k}_\perp|^2 + k_\parallel^2}\, ,   
\ee
where ${\bf k}_\perp$ is orthogonal to the direction picked out by the parallel ${\bf D}$ and ${\bf B}$ background fields. 

Recalling also that $A_0 = (\alpha\beta-\gamma^2)/\mathcal{H}^2$, we see that $A_0<0$ when $(\alpha\beta-\gamma^2)<0$ (the rBI and Pleba{\'n}ski cases) because $\mathcal{H}^2$ must be non-negative for real $\mathcal{H}$. We then have $v_g^2>1$ for any choice of ${\bf k}$ for which $v_g$ is real and $|{\bf k}_\perp|\ne0$.  For $\alpha\beta-\gamma^2>0$ we have $A_0>0$ but $A_0>1$ is still possible,  which would imply $v_g^2>1$ for a static background with  $(\alpha x+\beta y + \gamma z )<0$ and $|{\bf k}_\perp|\ne0$.  To avoid both possibilities for superluminal propagation we need precisely the convexity conditions \eqref{convex2}, which restrict the possible causal theories to BI and its extreme limits (the BB limit is excluded here as it has no static backgrounds).  In these cases $0\le A_0\le 1$, which ensures that $v_g^2\le1$. This extends a result of \cite{Bialynicki-Birula:1984daz} for BI to the extreme BI limits. As $A_0=0$ for the extreme BI limits, $v_g=1$; this is a result of \cite{Russo:2022qvz}, where it was also found that wave propagation in a static background is restricted to the direction 
determined by the background fields. 

We shall now generalise to a generic stationary background, assuming the convexity conditions hold. As we have seen, the dispersion relation \eqref{Unique} is the unique one for the entire three-parameter family of NLEDs defined by $\mathcal{H}= \sqrt{Q}$; it is 
\be\label{stat.rec}
\tilde\omega^2 = A_0 k_1^2 + \Phi(k_2,k_3)\, , 
\ee
where (we recall)
\be\label{Phi}
\tilde\omega = \omega - k_1\frac{p}{\mathcal{H}}\, , \qquad \Phi= Kk_2^2 + Ak_3^2 -2C k_2 k_3\, . 
\ee
From this we see that 
\be
\boldsymbol{v}_g \equiv \frac{d\omega}{d{\bf k}} = \mathcal{H}^{-1}\left(p + \frac{(\alpha\beta-\gamma^2) k_1}{\tilde\omega\mathcal{H}}\right) {\bf e}_1 +
\frac{(Kk_2-Ck_3)}{\tilde\omega} {\bf e}_2 + \frac{(Ak_3-Ck_2)}{\tilde\omega} {\bf e}_3\, .  
\ee
Using the fact that 
\be 
\begin{aligned}
(Kk_2 -Ck_3)^2 + (Ak_3-Ck_2)^2 &=  (A+K) \Phi - (AK-C^2)(k_2^2+k_3^2)\, \\
&= \Phi\left(1 - \frac{\left[p^2 -(\alpha\beta-\gamma^2)\right]}{\mathcal{H}^2} \right) - 
A_0\left(k_2^2+ k_3^2\right)\, , 
\end{aligned}
\ee
and the fact that 
\be
\frac{\Phi}{\tilde\omega^2} = 1- \left(\frac{A_0}{\tilde\omega^2}\right) k_1^2\, , 
\ee
we find that
\be\label{vg-stat}
v_g^2 = 1+ \frac{A_0}{\tilde\omega^2} \left[\omega^2 - |{\bf k}|^2\right]\, . 
\ee
The formula \eqref{vg}, derived for static backgrounds, can be recovered by using the fact that 
$\tilde\omega=\omega$ and $\omega^2= k_2^2 + A_0(k_1^2+k_3^2)$ when $p=0$.  

We see immediately from \eqref{vg-stat} that $v_g=1$ when $A_0=0$; i.e. for all three BI limits, and in any stationary homogeneous background.
What we do {\sl not} see from this equation is whether there are restrictions on the possible directions of wave propagation. We return to this issue in the next section. 

For $A_0>0$; i.e the BI theory,  we learn from eq. \eqref{vg-stat} that the group velocity is never superluminal iff the phase velocity is never superluminal. We know (from the above review) that the phase velocity is never superluminal in a static background, and this will remain true after a boost to a stationary background because $(\omega^2- |{\bf k}|^2)$ is Lorentz invariant. However, there may be stationary backgrounds that are not obtainable in this way; they are {\sl intrinsically stationary}.  We shall see in the next subsection that there are such backgrounds for BI: those for which ${\bf D}\cdot {\bf B}=0$ and $|{\bf D}|= |{\bf B}| = \sqrt{p}$. This 
means that the analysis of causality for BI in \cite{Bialynicki-Birula:1984daz} is not complete.  We complete it now.  

For an intrinsically-stationary BI backgrounds the BI dispersion relation \eqref{BI-wave} simplifies to 
\be
(\omega - k_1) \left[\omega + \left(\frac{T-p}{T+p}\right)k_1\right] = \frac{T}{(T+p)}\left( k_2^2+k_3^2\right) \,.
\ee
The two solutions are $\omega=\omega_\pm$, where
\be
\omega_\pm = \frac{1}{T+p}\left\{ p\,  k_1 \pm \sqrt{T^2 k_1^2 + T(T+p)(k_2^2+k_3^2)}\right\}\, . 
\ee
Assuming (without loss of generality) that $k_1>0$, 
then $\omega_+^2>\omega_-^2$ but
\be
(T+p)^2(\omega_+^2-|{\bf k}|^2)= - p \left(\sqrt{T}k_1-\sqrt{T k_1^2 + (T+p)(k_2^2+k_3^2)}\right)^2\leq 0\ ,
\ee
and hence, from \eqref{vg-stat},  $v_g\le 1$.

\subsection{Stress-Energy tensor redux}

The structure and properties of the stress-energy for a generic relativistic NLED defined by its Hamiltonian density were discussed in subsection \ref{subsec:stress}.
Now we specialize to the class of zero-birefringence NLEDs with $\mathcal{H}=\sqrt{Q}$; i.e. \eqref{quadclass}. 

We start with the BI case, for which 
\be
\mathcal{H} = \sqrt{p^2 + 2T(x+y) + T^2}\, ,  \qquad W=  \frac{p^2 + T(x+y)}{\mathcal{H}}
\ee
and hence 
\be
\begin{aligned}
2W- \mathcal{H} &=  \mathcal{H}^{-1} (p^2-T^2)\, , \\
W_\pm -\mathcal{H} &= - \mathcal{H}^{-1} \left\{ T^2 + T\left[(x+y) \mp  \sqrt{(x-y)^2 +z^2}\right] \right\}\, . 
\end{aligned}
\ee
These results yield the following stress-energy tensor for a generic static ($p=0$) background solution:
\be\label{stress-static}
\Theta_{\rm BI}^{\rm static} = \left(\begin{array}{cccc} \bar{\mathcal{H}} & 0 & 0 & 0 \\   0 &- T^2/\bar{\mathcal{H}} & 0 & 0 \\ 0 & 0 & - \bar{\mathcal{H}} & 0\\ 
0 & 0 & 0 &  - T^2/\bar{\mathcal{H}}  \end{array}\right) \, , \qquad \left(\bar{\mathcal{H}} = \sqrt{T^2+2T(x+y)}\right).
\ee
This is the stress-energy tensor of a tensile  optical medium (since the pressures are negative). It is anisotropic  due the background vector fields in the ${\bf e}_2$ direction, and when these fields are absent it reduces to $\Theta=-T \eta$, where $\eta$ is the Minkowski metric. 
Recalling \eqref{staticA0} and \eqref{static-disp},  we see that the  dispersion relation for wave propagation in a static background is 
\be\label{BIdr}
\omega^2 = k^2_\parallel + \left(\frac{T}{\bar{\mathcal{H}}}\right)^2 |{\bf k}_\perp |^2\, ,  
\ee
where ${\bf k}_\perp$ is the projection of ${\bf k}$ onto the plane orthogonal to ${\bf e}_2$.  The anisotropy of 
the stress-energy tensor is reflected in this dispersion relation,

Let us now consider the eBI case. We have
\be
\mathcal{H} = \sqrt{p^2 + 2Tx}\, , \qquad W= \frac{p^2+ Tx}{\mathcal{H}}\, , 
\ee
and hence 
\be
2W- \mathcal{H} = \frac{p^2}{\mathcal{H}}\, , \qquad W_- -\mathcal{H} = -\frac{2T x}{\mathcal{H}} \, , \qquad W_+-\mathcal{H} =0\, . 
\ee
For a generic static ($p=0$) background we now find that the background stress-energy tensor is 
\be\label{stress-eBI}
\Theta_{\rm eBI}^{static} = \left(\begin{array}{cccc} \bar{\mathcal{H}} & 0 & 0 & 0 \\   0 &0  & 0 & 0 \\ 0 & 0 & - \bar{\mathcal{H}} & 0\\ 
0 & 0 & 0 &  0 \end{array}\right) \, , \qquad \left(\bar{\mathcal{H}} = \sqrt{2Tx}\right). 
\ee
The two pressures in the plane orthogonal to the background vector fields are now zero. We might expect to find that 
waves cannot propagate in directions with zero pressure, in which case we would predict that eBI waves in a static background
can propagate only in the direction of the background vector fields. Furthermore, we could predict that these waves will be lightlike
since the one non-zero pressure equals the energy density. These predictions, based on the background stress-energy tensor are 
precisely what was found in \cite{Russo:2022qvz} from a direct computation: the dispersion relation is now $\omega^2 = k^2_\parallel$. 

It is important to appreciate here that any contribution to the stress-energy tensor arising from the addition of a constant term to the 
Hamiltonian density is being excluded because it has {\sl no effect on the NLED field equations}. We could subtract $\bar{\mathcal{H}}$ from the BI Hamiltonian in order to have zero vacuum energy; this would lead to a zero stress-energy tensor for the NLED vacuum ($x=y=0$). This 
might sound reasonable but only the {\sl intrinsic} value of the vacuum energy ($T$ for BI) has any effect on the NLED physics. A change in $T$ 
changes the dispersion relation  \eqref{BIdr} but a change in the constant energy density added to $\mathcal{H}_{\rm BI}$ to normalize the vacuum energy to some preferred value changes nothing (except the cosmological constant in a gravitational context). 

We now turn to stationary backgrounds. Since these are generically boosts of static backgrounds, we need to know which 
stationary backgrounds are not of this type, and the conclusion of subsection \ref{subsec:stress} was that $W^2=p^2$ for these "intrinsically stationary" 
backgrounds (we do not assume for the moment that $W>0$).  Using \eqref{firstd} we have
\be
W= \frac{ p^2 + (\alpha x+\beta y +\gamma z)}{\mathcal{H}}\, , 
\ee
which yields
\be
W^2-p^2 = \mathcal{H}^{-2} \left\{ (\alpha x+\beta y +\gamma z)^2 - (\alpha\beta-\gamma^2)p^2\right\}\, . 
\ee
We see that $W^2=p^2$  when
\be\label{W=p}
\boxed{ (\alpha x+\beta y + \gamma z)^2 = (\alpha\beta-\gamma^2)p^2 }
\ee
Let us examine this for various special cases:
\begin{itemize}

\item $(\alpha\beta-\gamma^2)<0$; i.e. rBI or its Pleba{\'n}ski limit. There are no non-vacuum configurations satisfying \eqref{W=p}. In addition, these cases allow $W<0$
since either $\alpha<0$ or $\beta<0$. As we already seen, these are unphysical theories because they allow superluminal propagation.  

\item $(\alpha\beta-\gamma^2)>0$; i.e. BI. We may choose  $\alpha=\beta = T>0$ and $\gamma=0$, in which case \eqref{W=p} reduces to 
\be
(x-y)^2 + z^2 =0 \qquad \Rightarrow \quad x=y\, , \quad z=0\, . 
\ee
In other words, a homogeneous stationary background in which ${\bf D}$ is orthogonal to ${\bf B}$, and $|{\bf D}|= |{\bf B}| = \sqrt{p}$, is {\sl intrinsically} stationary. For this background, we have
\be\label{backHBI}
\mathcal{H}_{BI} = p+T\, , 
\ee
which yields the stress-energy tensor
\be\label{ThetaBI}
\Theta_{\rm BI} = \left(\begin{array}{cccc} p+T & p & 0 & 0 \\ p  & p-T & 0 & 0 \\ 0 & 0 & -T  & 0\\ 0 & 0 & 0 & -T \end{array}\right) \, .   
\ee
We again have planar isotropy but the plane is now that spanned by the background vector fields. 

\item $\alpha\beta-\gamma^2=0$ and $\alpha+\beta =T>0$. These are the extreme BI limits. We may solve the 
constraint on parameters by setting
\be
(\alpha,\beta,\gamma) = T(\cos^2\theta, \sin^2\theta, \sin\theta\cos\theta)\, ,  
\ee
for constant $T$ and angle $\theta$, in which case
\be
\alpha x +\beta y + \gamma z = T |{\bf D}_\theta|^2\, , 
\ee
where
\be\label{def-Dtheta}
\begin{aligned}
{\bf D}_\theta &=\ \ \cos\theta\, {\bf D} + \sin\theta\, {\bf B}\, ,
\\
{\bf B}_\theta &= - \sin\theta\, {\bf D} + \cos\theta\, {\bf B} \, .
\end{aligned}
\ee
The equation \eqref{W=p} has no non-vacuum solution, so all stationary background solutions are Lorentz boosts of a static background. The background stress-energy tensor is therefore a Lorentz boost of \eqref{stress-eBI} with 
$\bar{\mathcal{H}} \to \mathcal{H}_\theta$:
\be\label{Htheta1}
\Theta_{\theta} = \left(\begin{array}{cccc} \mathcal{H}_\theta & p & 0 & 0 \\   p &\frac{p^2}{\mathcal{H}_\theta}  & 0 & 0 \\ 0 & 0 & - \mathcal{H}_\theta & 0\\ 
0 & 0 & 0 &  0 \end{array}\right) \, , \qquad \left(\mathcal{H}_\theta = \sqrt{p^2 + T|{\bf D}_\theta|^2}\right). 
\ee
Notice that both eBI and meBI are now unified in a way that makes it manifest that one is the electromagnetic dual of the other. 
The tension is now zero only in one direction, which is the direction of ${\bf D}_\theta$, so we can expect wave propagation in all directions 
orthogonal to ${\bf D}_\theta$. We shall verify this prediction in the next section. 

\item $\alpha=\beta=\gamma=0$. Now $\mathcal{H}=p$. This is BB and the (intrinsically stationary) background has the 
stress-energy tensor 
\be
\Theta_{\rm BI} = \left(\begin{array}{cccc} p & p & 0 & 0 \\ p  & p & 0 & 0 \\ 0 & 0 & 0  & 0\\ 0 & 0 & 0 & 0 \end{array}\right) \, .   
\ee
The pressure is now zero 
in the plane orthogonal to ${\bf p}$, so we expect wave propagation to be possible only in this direction. 
Again, we shall see in the following section that this is true, 

\end{itemize}

We have not yet mentioned Maxwell electrodynamics. This can be viewed as the infinite tension limit of BI, although this limit can be 
taken only if the ``intrinsic'' vacuum energy density $T$ is first subtracted. However, this  $-T$ added to $\mathcal{H}_{\rm BI}$ has no effect
on the NLED field equations, and hence no effect on the dispersion relations, so the $T\to\infty$ limit of \eqref{BIdr} should yield the usual 
Maxwell dispersion relation $\omega^2= |{\bf k}|^2$, and it does.

\section{The ``extreme'' limits of Born-Infeld}

We now focus on the ``extreme'' limits of BI, with the Hamiltonian density $\mathcal{H}_\theta$ of \eqref{Htheta1}; equivalently 
\be\label{Htheta}
\mathcal{H}_\theta = \sqrt{|{\bf D}\times {\bf B}|^2 + T|\cos\theta\, {\bf D} + \sin\theta\, {\bf B}|^2}\,  \qquad (T>0).
\ee
The choices $\sin\theta=0$ and $\cos\theta=0$ correspond, respectively,  to the electric and magnetic  ``extreme'' limits of Born-Infeld introduced in 
\cite{Russo:2022qvz}: eBI  and meBI. 

As noted in section \ref{sec:quad}, a shift  ${\bf D} \to {\bf D} - \epsilon {\bf B}$ adds a 
total derivative to the phase-space Lagrangian density $\widetilde{\mathcal{L}}$, so
Hamiltonians related by such redefinitions are therefore equivalent. For $\cos\theta\neq 0$ 
we can use this freedom to eliminate the $\sin\theta\, {\bf B}$ term. The ($\theta=0$) eBI Hamiltonian can then be recovered by a rescaling of $T$. In contrast, for  $\cos\theta =0$, the Hamiltonian is unchanged by a shift in ${\bf D}$. These features will be explicit in the Lagrangian densities given at the end of this section.

\subsection{Hamiltonian field equations}

In terms of the fields $({\bf D}_\theta, {\bf B}_\theta)$ defined in \eqref{def-Dtheta},  
the Hamiltonian field equations for the extreme limits of BI simplify to 
\be\label{Extreme-feqs}
\begin{aligned}
\dot{\bf D}_{\theta}  =& \boldsymbol{\nabla} \times \left[{\bf n}_T \times {\bf D}_\theta \right]\, ,
 \qquad \qquad \ \boldsymbol{\nabla} \cdot {\bf D}_\theta=0\, , \\
\dot{\bf B}_\theta =& \boldsymbol{\nabla} \times \left[ {\bf n}_T\times {\bf B}_\theta
- \frac{T}{\mathcal{H}} {\bf D}_\theta \right]\, , \quad \boldsymbol{\nabla} \cdot {\bf B}_\theta=0\, , 
\end{aligned}
\ee
where
\be
{\bf n}_T = \mathcal{H}^{-1}_\theta {\bf D}_\theta\times {\bf B}_\theta \ . 
\ee
These equations are just the macroscopic Maxwell equations for $({\bf D},{\bf B})$ with 
\be
{\bf D}_\theta  + i{\bf B}_\theta = e^{i\theta} \left({\bf D} + i{\bf B}\right)\, ,  
\ee
i.e. a duality rotation, but this is a field definition that can be made only in the field equations, and not in the phase-space action because ${\bf D}$ is not divergence-free ``off-shell''.
Notice that 
\be
|{\bf n}_T|^2 = 1- \frac{T|{\bf D}_\theta|^2}{\mathcal{H}^2}\le 1\, , 
\ee
with equality only for $T=0$, for which the  equations \eqref{Extreme-feqs} reduce to the 
BB field equations \cite{Bialynicki-Birula:1984daz}:
\be\label{BBeqs}
\begin{aligned}
\dot{\bf D}  =& \boldsymbol{\nabla} \times \left[{\bf n} \times {\bf D}\right]\, ,
 \qquad \qquad \ \boldsymbol{\nabla} \cdot {\bf D}=0\, , \\
\dot{\bf B} =& \boldsymbol{\nabla} \times \left[ {\bf n}\times {\bf B} \right]\, , \  \qquad \qquad \boldsymbol{\nabla} \cdot {\bf B}=0\, , 
\end{aligned}
\ee
where ${\bf n}$ is now a unit vector field:
$$
{\bf n}= \frac{{\bf D}\times {\bf B} }{|{\bf D}\times {\bf B} |} \ .
$$ 
It is instructive to look for simple solutions of the equations \eqref{Extreme-feqs}
or \eqref{BBeqs}. 

\begin{itemize}
    
\item \boxed{{\rm Extreme\  BI}.} For time-independent configurations
satisfying $|{\bf D}\times {\bf B}|=0$ we have 
$\mathcal{H}_\theta = \sqrt{T} |{\bf D}_\theta|$, and the field equations \eqref{Extreme-feqs} reduce to
\be
\boldsymbol{\nabla}\times\left(\frac{{\bf D}_\theta}{|{\bf D}_\theta|}\right) =0\, , \qquad 
\boldsymbol{\nabla}\cdot {\bf D}= 
\boldsymbol{\nabla}\cdot {\bf B}=0\, , 
\ee
which are solved for any $\theta$ by
\be
{\bf B}= B(x^1,x^2)\, {\bf e}_3\ ,\qquad 
{\bf D}=D(x^1,x^2)\, {\bf e}_3\ .
\ee
In the Lagrangian formulation, discussed at the end of this section, the electric field ${\bf E}$ for this solution is
\be
{\bf E}= \left\{\begin{array}{ccc} \sqrt{T} {\bf e}_3 && {\rm eBI} \\ 
{\bf 0} && {\rm meBI} \end{array}\right\}\, , 
\ee
which illustrates the fact that many functions ${\bf D}$ correspond 
to the same ${\bf E}$, which is a consequence of a zero-determinant Hessian matrix for the Hamiltonian density. It also shows that a non-zero $\boldsymbol{\nabla} \times {\bf B}$ is compatible with a constant electric field, in contrast to the Maxwell theory where (since ${\bf D}={\bf E}$ and ${\bf H}={\bf B}$, and hence $\dot {\bf E} = \boldsymbol{\nabla} \times {\bf B}$) it implies a time-dependent electric field. 
\bigskip

\item \boxed{\rm BB\ electrodynamics.} For any constant uniform unit vector ${\bf n}$ the field equations \eqref{BBeqs} reduce to the linear equations
\be\label{ncons}
\begin{aligned}
\dot{\bf D}  =& -\left({\bf n} \cdot \boldsymbol{\nabla}\right) {\bf D}\, ,
 \qquad \qquad \ \boldsymbol{\nabla} \cdot {\bf D}=0\, , \\
\dot{\bf B} =& -\left({\bf n} \cdot \boldsymbol{\nabla}\right) {\bf B}\, \, , \  \qquad \qquad \boldsymbol{\nabla} \cdot {\bf B}=0\, . 
\end{aligned}
\ee
Choosing ${\bf n}={\bf e_1}$ we have the solution
\be
{\bf B}=B(t-x^1,x^3) \ {\bf e_2}\ ,  \qquad 
{\bf D}=D(t-x^1,x^2)\ {\bf e_3}\ ,
\ee
provided that $DB\ne 0$ (otherwise ${\bf n}$ is not defined). This is a wave in the 1-direction with non-trivial profile; its Fourier components are solutions of the linearized BB equations with $\omega = k_1$, where linearization is about the solution with non-zero constant modes for $D$ and $B$. 
\end{itemize}

In general, the Hamiltonian field equations imply the continuity conditions
\be
\partial_\mu T^{\mu\nu}=0 \qquad  (\mu,\nu =0,1,2,3). 
\ee
This remains true in the $T\to 0$ limit, for which we find the BB stress-energy tensor:
\be
T^{\mu\nu}_{\rm BB} =  n^\mu n^\nu \mathcal{H}_{\rm BB} \ , 
\qquad n^\mu = (1, {\bf n} := {\bf p}/p)\, .    
\ee
Since ${\bf n}$ is now a unit 3-vector field, BB electrodynamics is a dynamical theory for a null fluid \cite{Bialynicki-Birula:1992rcm}.


\subsection{Wave propagation} 

The dispersion relation of \eqref{stat.rec} simplifies considerably for $\mathcal{H}_\theta$  because 
the quadratic form $\Phi$ is now a perfect square:
\be
\Phi =  \frac{T^2}{\mathcal{H}_\theta^2} \left[(D_\theta)_2k_2 + \sin\theta B_3 k_3\right]^2\, .  
\ee
Consequently, the quadratic dispersion relation \eqref{Unique} degenerates to a pair of linear relations; written for our particular choice of axes they are
\be
\omega = \mathcal{H}_\theta^{-1}\left\{ pk_1  \pm T\left[(D_\theta)_2 k_2  + \sin\theta B_3 k_3\right]\right\}\, . 
\ee
For an arbitrary choice of axes these relations take the form
\be
\omega = \boldsymbol{v}_g^{(\pm)} \cdot {\bf k}\, , 
\ee
where\footnote{We are omitting the bars on background fields here.}
\be\label{vgpm}
\boldsymbol{v}_g^{(\pm)} = \mathcal{H}_\theta^{-1} \left\{{\bf p}  \pm T {\bf D}_\theta \right\}\, . 
\ee
Using \eqref{Htheta} and the fact that ${\bf p}\cdot {\bf D}_\theta =0$, we confirm that $v_g=1$ for either choice of sign, and this is also the phase velocity.   
In the $T\to0$ limit $\mathcal{H}_\theta \to \mathcal{H}_{\rm BB}=p$, and $\boldsymbol{v}_g^{(\pm)} = {\bf p}/p = {\bf n}$.
Thus, all the limits of BI have the property that plane-wave perturbations of a stationary homogeneous background are lightlike with a pair of linear dispersion relations, which coincide
in the BB limit. 

As expected from our discussion of the stress-energy tensor at the end of the 
previous section, there is no propagation in directions orthogonal to the 
plane spanned by the two orthogonal 3-vectors $({\bf p},{\bf D}_\theta)$. 
The specific direction of propagation in this plane depends on the ratio of 
${\bf k} \cdot{\bf p}$ to ${\bf k}\cdot {\bf D}_\theta$. In the special case
that $p=0$ we recover the result of \cite{Russo:2022qvz} that propagation is
necessarily parallel to ${\bf D}_\theta$. 

Notice that there is no special case for which $|{\bf D}_\theta|=0$ because then $p=0$ too
and $\mathcal{H}=0$. However, in the $T\to0$ limit we get the BB theory, for which any 
constant uniform non-vacuum background has non-zero $p$ because 
$\mathcal{H}_{\rm BB}=p$; in this case propagation is necessarily in the direction of 
${\bf p}$. This too can be understood from the stress-energy tensor because the pressure of the background medium is zero in all directions orthogonal to ${\bf p}$, hence the ``photon dust'' interpretation of \cite{Bialynicki-Birula:1992rcm}.

\subsection{Lagrangian formulation}

To pass to the Lagrangian formulation we return to the Hamiltonian density in the 
general form of \eqref{quadclass}, but now with $\alpha\beta\ge0$ and $\gamma=\pm \sqrt{\alpha\beta}$. We then introduce the electric 
field via the definition 
\be\label{defE}
{\bf E}= \frac{\partial \mathcal{H} }{\partial \DD} =\frac{1}{ \mathcal{H}}\left[ \left(|{\bf B}|^2+\alpha\right){\bf D} + 
\left(\gamma-{\bf D} \cdot {\bf B}\right) {\bf B} \right]\, ,  
\ee 
from which we deduce that
\be\label{P.ext}
P := \EE\cdot \BB= \frac{1}{ \mathcal{H}}\left( \alpha \DD\cdot \BB + \gamma  |\BB|^2 \right) 
\ee 
and also that 
\be\label{S}
2S:= |{\bf E}|^2 -|{\bf B}|^2 = \alpha  - \frac{1}{\mathcal{H}^2}
\left[\sqrt{\alpha} ({\bf D}\cdot {\bf B}) \pm \sqrt{\beta}\, |{\bf B}|^2\right]^2\, . 
\ee 
where the sign choice corresponds to the sign choice made for $\gamma$. 
A consequence of these relations is the identity 
\be\label{canonC}
\alpha^2 -2\alpha S -P^2 \equiv 0\, .  
\ee
This confirms that there cannot be a unique solution of \eqref{defE}
for ${\bf D}$, as expected from the fact that the Hessian matrix of $\mathcal{H}$ as a function of ${\bf D}$ has a zero eigenvalue. 

Let us attempt to obtain a `canonical' Lagrangian density in the standard way:
\be\label{formula}
\begin{aligned}
\mathcal{L}_{({\rm can})} =&\  {\bf D} \cdot {\bf E} - \mathcal{H}\\
=&\  \frac{1}{\mathcal{H}}\left[ |{\bf D} \times {\bf B}|^2 + (\alpha |{\bf D}|^2 
+ \gamma ({\bf D}\cdot {\bf B})] - \mathcal{H}\right] \\
=&\  - \frac{1}{\mathcal{H}}\left[ \beta|{\bf B}|^2 
+ \gamma {\bf D}\cdot {\bf B}\right] \, .
\end{aligned}
\ee 
To proceed, we shall now consider in turn the cases for which $\alpha\ne0$ and $\alpha=0$:

\begin{itemize}

\item $\boxed{\alpha\ne0.}$ By comparing \eqref{P.ext} with \eqref{formula} we see that 
\be
\mathcal{L}_{({\rm can})} = -(\gamma/\alpha) P\, . 
\ee
Imposing the Lagrangian constraint with Lagrange multiplier $\lambda$ yields
\be
\mathcal{L} = -(\gamma/\alpha) P  - \lambda\left[ \alpha^2 - 2\alpha S -P^2\right]\, . 
\ee
Setting $(\alpha,\beta)=(T,0)$ (and hence $\gamma=0$),  we 
recover the eBI Lagrangian density found in \cite{Russo:2022qvz}:
\be\label{eBI-L}
\mathcal{L}_{\rm eBI} = -\lambda(T^2 - 2T S -P^2)\, . 
\ee
For $\beta\ne 0$ the only difference is the addition of a total derivative.

\item $\boxed{\alpha=0.}$ In this case $\gamma=0$ and we may set $\beta=T$. Eq. \eqref{S} now simplifies to 
\be
2TS = - \left(\frac{T|{\bf B}|^2}{\mathcal{H}}\right)^2 = - \mathcal{L}^2_{({\rm can})}\, ,  
\ee
which requires $S\le0$ for $T>0$, and yields
\be
\mathcal{L}_{({\rm can})} =- \sqrt{-2TS}\, . 
\ee
In addition, \eqref{P.ext} becomes the Lagrangian constraint $P\equiv0$, which is
consistent with, and replaces, \eqref{canonC}. Imposing this constraint we recover 
the meBI Lagrangian density found in \cite{Russo:2022qvz}:
\be
\mathcal{L}_{\rm meBI} = - \sqrt{-2TS} - \lambda P\, . 
\ee 
Note that in this case the addition of a total derivative term
proportional to $P$ has no effect since it can be removed by a redefinition of $\lambda$.

\end{itemize} 
It is instructive to check these results by an inverse Legendre transform. 

\begin{itemize}

\item $\boxed{\rm eBI}$ Starting from \eqref{eBI-L} we define
\be
{\bf D} := \frac{\partial \mathcal{L}_{\rm eBI}}{\partial {\bf E}} 
= \lambda \left(T{\bf E} + P{\bf B}\right)\, ,  
\ee
from which we deduce that 
\be\label{Ededuce}
{\bf E} = \frac{ (T+ |{\bf B}|^2) {\bf D} - ({\bf D}\cdot{\bf B}) {\bf B}}{ \lambda T (T+|{\bf B}|^2)}\, \qquad \left[\Rightarrow\  P= \frac{{\bf D}\cdot {\bf B}}{\lambda \left(T+|{\bf B}|^2\right)}\right]\, , 
\ee
and hence\footnote{This corrects eq.(4.8) of \cite{Russo:2022qvz} in which the $-TP^2$ term is missing.}
\be
T^2|{\bf E}|^2 = \frac{|{\bf D}\times {\bf B}|^2 + T|{\bf D}|^2}{\lambda^2(T+|{\bf B}|^2)} -TP^2\, . 
\ee
Using these relations, we find that 
\be\label{Hcan}
\mathcal{H} := {\bf D}\cdot {\bf E} - \mathcal{L}_{\rm eBI} =
\frac12\left\{ \ell^{-1} \left[|{\bf D}\times {\bf B}|^2 + T|{\bf D}|^2\right] + \ell\right\}\, , 
\ee
where 
\be\label{ell-lam}
\ell= \lambda T(T+|{\bf B}|^2)\, .   
\ee
Notice that the Lagrange multiplier $\lambda$ has now become the auxiliary field $\ell$, with field equation 
\be\label{aux-soln}
\ell = \sqrt{|{\bf D}\times{\bf B}|^2 + T|{\bf D}|^2} = \mathcal{H}_{\rm eBI}\, . 
\ee
Back-substitution in \eqref{Hcan} yields $\mathcal{H}= \mathcal{H}_{\rm eBI}$.

\item $\boxed{\rm meBI}$ It is convenient to start from the following equivalent Lagrangian density involving an 
auxiliary field $\ell$:
\be\label{Lprime}
\mathcal{L}'_{\rm meBI} = \frac12\left\{ \ell(2TS) - \ell^{-1}\right\} - \lambda P\, .
\ee
Now we have
\be
{\bf D} := \frac{\partial \mathcal{L}'_{\rm meBI}}{\partial {\bf E}} = \ell T {\bf E} -\lambda {\bf B}\, , 
\ee
and hence 
\be\label{E-meBI}
{\bf E} = (\ell T)^{-1}\left({\bf D} +\lambda {\bf B}\right)\, . 
\ee
Using this to eliminate ${\bf E}$ we find that
\be
\mathcal{H}' = {\bf D}\cdot {\bf E} - \mathcal{L}'_{\rm meBI} = \frac12\left\{ e^{-1} \mathcal{H}_{\rm meBI}^2 + e\right\} + \frac{1}{2e}\left[\lambda |{\bf B}|^2 + {\bf D}\cdot{\bf B} \right]^2\, , 
\ee
where
\be
e= \ell T |{\bf B|}^2\, . 
\ee
The Lagrange multiplier $\lambda$ has  become an additional auxiliary field $e$. 
Upon elimination of both auxiliary fields, we get
\be
\lambda =-\frac{{\bf D}\cdot{\bf B}}{|{\bf B}|^2}\ ,
\qquad e= \mathcal{H}_{\rm meBI}\ ,
\ee
so that $\mathcal{H}' \to  \mathcal{H}_{\rm meBI}$. 

\end{itemize}

It is evident from these results that the Lagrangian formulations of the two extreme limits of BI greatly obscures 
the fact that they are related by discrete electic/magnetic duality.

\section{Summary and Outlook}

This paper is a sequel to an earlier one in which the issue of birefringence in nonlinear electrodynamics (NLED) was revisited with the aim of finding a complete list of those theories for which constant uniform electromagnetic backgrounds provide a homogeneous optical medium {\sl without} birefringence \cite{Russo:2022qvz}. It was well-known that Born-Infeld (BI) was the unique zero-birefringence NLED given certain assumptions (the simplest being the assumption of a weak-field limit) but another ``pathological'' case found by Pleba{\'n}ski \cite{Plebanski:1970zz} had been frequently mentioned in the literature. Were there more cases to be found? 

By following a systematic method due to Boillat \cite{Boillat:1970gw},  another ``reverse-Born-Infeld'' (rBI) was found in \cite{Russo:2022qvz},  and a solution of the  ``Boillat equations'' at the parameter boundary between BI and rBI was shown to lead to a Lagrangian constraint rather than a Lagrangian. It was argued that the `non-standard' Lagrangian found by imposing this constraint with a Lagrange multiplier would define yet another zero-birefringence NLED, which was called ``extreme-Born-Infeld'' (eBI); one of the arguments was that the eBI Hamiltonian was not only `standard'  but also a (non-conformal) scaling limit of the BI Hamiltonian. The electromagnetic duality of the BI Hamiltonian is lost in this scaling limit, and this implies the existence of a `dual' scaling limit of BI with a reversed role for electric and magnetic fields; this was called ``magnetic-extreme-BI'' (meBI) and it was shown that the meBI Lagrangian is also `non-standard' because of a Lagrangian constraint. 

As we have pointed out here, both ``extreme''  limits of BI and the conformal strong-field BB limit are solutions of the Boillat equations
for zero-birefringence once one allows solutions involving Lagrangian constraints. Including them yields a list of six distinct zero-birefringence NLEDs, but as the derivation of the Boillat equations starts from a standard Lagrangian density without constraints, 
the birefringence status of the three BI-limits on the list is not settled by this observation. However, it {\sl is} settled by our 
Hamiltonian birefringence analysis (and, for the BB case, the previous Hamiltonian analysis of \cite{Bandos:2023yat}). 

An issue that we have passed over is whether our list of six zero-birefringence NLEDs is complete. To address this issue 
one must first decide what ``completeness'' means; i.e. within what class of NLEDs. We could choose the class for which 
the Hamiltonian density $\mathcal{H}({\bf D},{\bf B})$ is any (sufficiently differentiable) function but this would include
many NLEDs that are not Lorentz invariant. We have chosen to impose Lorentz invariance (but not electromagnetic 
duality invariance, which would eliminate all but BI and BB from our list). It is also reasonable to require $\mathcal{H}$ to be a convex function of the electric displacement vector field ${\bf D}$,  because this guarantees the existence of an equivalent Lagrangian, 
and because it is required by causality. 

Imposing ``strict convexity'' would eliminate all but BI from our list, and ensure its unique zero-birefringence status; 
this is because ``strict convexity'' ensures the existence of a `standard' manifestly Lorentz invariant Lagrangian without 
constraints, and in this context the Boillat equations both apply and allow only BI. The BI limits would be excluded as 
the Hessians of their Hamiltonian densities have zero eigenvalues, but there is no good physical reason to exclude them.
Any other zero-birefringent NLEDs within this larger class (satisfying the weaker convexity condition) must have a 
Hamiltonian density for which the Hessian matrix also has a zero determinant, but in some different way. 
We have not excluded this possibility but we think it unlikely for two reasons. One is the fact that our ``list of six''
can all be found as solutions of the Boillat equations, which suggests that these equations are valid within the larger NLED 
class. The other is the fact that the Hamiltonian densities of the ``list of six'' all have the same very simple form, which suggests 
that we have found the exact solution of the zero-birefringence conditions found here, but that some further insight is needed to prove it. This is plausible because it is also far from obvious how to solve the standard {\sl Lagrangian} zero-birefringence conditions; it only becomes obvious when they are rewritten as the ``Boillat equations''. 

Another aim of this paper has been to explore further the novel ``extreme'' limits of Born-Infeld. It was shown in \cite{Russo:2022qvz} that perturbations of static homogeneous backgrounds always propagate at light speed and only in a direction that is (anti)parallel to the background fields. Here we have provided a physical explanation for this unusual feature: the optical medium provided by the background has zero pressure in the directions in which the perturbations cannot propagate. The same explanation applies to non-static but stationary backgrounds, but now the medium has zero pressure in only one direction, and propagation is possible only in the plane orthogonal to it.  

This explanation of the unusual features of the limits of Born-Infeld relies on the fact that we exclude from the Hamiltonian density any 
constant term, such as that conventionally included to normalize the vacuum energy to zero. Whenever the resulting ``intrinsic'' 
stress-energy tensor has zero pressure in a given direction there is no wave propagation in this direction. While this is an expected feature 
in the context of a conventional optical medium, it leads to the conclusion that the Born-Infeld vacuum is an optical medium with 
tension equal to the Born constant $T$, but this vacuum is just Minkowski spacetime. The interpretation of Minkowski spacetime as a 
tensile medium becomes natural in the context of D3-brane dynamics, which reduces on a planar static brane (and omitting fermionic fields) 
to Born-Infeld,  and it supports the interpretation put forward in \cite{Townsend:2021wrs} of the D3-brane as an electromagnetic aether 
consistent with relativity. 

A concomitant feature of the restrictions on directions of propagation is the linearization of the quadratic dispersion relation, 
which becomes a pair of distinct linear dispersion relations for the extreme limits of BI, which coincide in the further conformal 
limit to BB. We have also confirmed, by a more unified calculation, the Lagrangian formulations of eBI and meBI found in 
\cite{Russo:2022qvz}.  We have just alluded to the significant role of Born-Infeld theory in the dynamics of D-branes in string theory, 
and we expect the limits of Born-Infeld to also play a role. A possible role for the conformal strong field BB limit has been proposed by two
of us \cite{Mezincescu:2019vxk}, but any role for the non-conformal ``extreme'' limits will likely be very different; 
one string-like feature  is that wave propagation in a static background is effectively reduced to left and right movers in one space dimension.

\section*{Acknowledgements}
PKT has been partially supported by the STFC consolidated grant ST/T000694/1. JGR acknowledges financial support from grants 2021-SGR-249 (Generalitat de Catalunya) and MINECO  PID2019-105614GB-C21.


\providecommand{\href}[2]{#2}\begingroup\raggedright\endgroup


\end{document}